\documentclass[preprint,aps,amsmath,amssymb,nofootinbib,superscriptaddress,tightenlines]{revtex4}

\usepackage{graphicx}
\usepackage{dcolumn}
\usepackage{bm}
\usepackage{verbatim}
\usepackage{amsmath}
\usepackage{amsfonts}
\usepackage{amssymb}

\makeatletter

\renewcommand{\p@subsection}{}

\renewcommand{\p@subsubsection}{}

\makeatother

\newcommand{\etal}{{\it et~al.} }
\newcommand{\sfrac}[2]{{\textstyle\frac{#1}{#2}}}

\begin{document}

\title{Cosmological Constraints on Decaying Dark Matter}

\author{Santiago~De~Lope~Amigo}%
\email{slamigo@physics.utoronto.ca}%
\affiliation{Department of Physics, University of Toronto, 60 St.
George Street, Toronto, Ontario M5S 1A7, Canada.}

\author{William~Man-Yin~Cheung}%
\email{mycheung@physics.utoronto.ca}%
\affiliation{Department of Physics, University of Toronto, 60 St.
George Street, Toronto, Ontario M5S 1A7, Canada.}

\author{Zhiqi~Huang}%
\email{zqhuang@astro.utoronto.ca}%
\affiliation{Department of Astronomy \& Astrophysics, University of Toronto, \\50 St.
George Street, Toronto, Ontario M5S 3H4, Canada.}
\affiliation{Canadian Institute for Theoretical Astrophysics, University of
Toronto,\\60 St. George Street, Toronto, Ontario M5S 3H8, Canada.}

\author{Siew-Phang~Ng}%
\email{spng@physics.utoronto.ca}%
\affiliation{Department of Physics, University of Toronto, 60 St.
George Street, Toronto, Ontario M5S 1A7, Canada.}

\date{\today}

\begin{abstract}
We present a complete analysis of the cosmological constraints on decaying dark matter. Previous analyses have used the cosmic microwave background and Type Ia supernova. We have updated them  with the latest data as well as extended the analysis with the inclusion of Lyman-$\alpha$ forest, large scale structure and weak lensing observations. Astrophysical constraints are not considered in the present paper. The bounds on the lifetime of decaying dark matter are dominated by either the late-time integrated Sachs-Wolfe effect for the scenario with weak reionization, or CMB polarization observations when there is significant reionization. For the respective scenarios, the lifetimes for decaying dark matter are $\Gamma^{-1} \gtrsim 100$ Gyr  and $ (f \Gamma) ^{-1} \gtrsim 5.3 \times 10^8$ Gyr (at 95.4\% confidence level), where the phenomenological parameter $f$ is the fraction of the decay energy deposited in baryonic gas. This allows us to constrain particle physics models with dark matter candidates through investigation of dark matter decays into Standard Model particles via effective operators. For decaying dark matter of $\sim 100$ GeV mass, we found that the size of the coupling constant in the effective dimension-4 operators responsible for dark matter decay has to generically be $ \lesssim 10^{-22}$. We have also explored the implications of our analysis for representative models in theories of gauge-mediated supersymmetry breaking, minimal supergravity and little Higgs.

\end{abstract}

\maketitle

\section{Introduction}

The past decade and a half has seen tremendous progress in the field of cosmology. The numerous experiments and the plethora of accumulated data have elevated cosmology into a precision science. With the knowledge we have gained, a remarkably consistent consensus known as the standard model of cosmology has emerged from these attempts to understand the nature of the universe. The picture that we have presently is that of a universe that is composed of 74\% dark energy, 22\% dark matter (DM) and 4\% baryonic matter \cite{Hinshaw2008, Komatsu2008}. Despite its extraordinary success in explaining a variety of diverse observations, fundamental questions do remain. Arguably the most vexatious is the question, ``What are these dark components of the universe?" Despite the fact that it makes up 96\% of the universe, we have so far been unable to say definitively what they are. Perhaps the most compelling ideas to resolve this question arise from particle physics. Dark energy is commonly attributed to the vacuum energy while the identity of dark matter is hypothesized to be one of the new particles in theories that  extend the Standard Model of particle physics. This confluence and cross-fertilization of ideas from two major fields of scientific endeavor promises to herald an exciting new era in understanding of the universe. With the resumption of operation of the Large Hadron Collider at CERN, we are possibly months away from collider detection, albeit indirectly through missing energy signatures, of dark matter particles.

Yet this avenue too is plagued with troubling questions that need to be addressed if we are to take the idea seriously that some particle from an extension of the Standard Model is indeed the elusive yet ubiquitous dark matter of the universe. Most worryingly, to ensure the presence of a dark matter candidate in a number of beyond the Standard Model theories of particle physics, it is often necessary to impose global symmetries. For instance, we have the T-parity in Little Higgs \cite{Schmaltz:2005ky} and R-parity in supersymmetry \cite{Martin:1997ns}. In the limit where the global symmetries are exact, the lightest particle carrying such a global charge would be stable from decay to lighter particles that do not possess such a charge. It is this point that could potentially destroy this promising marriage of ideas from cosmology and particle physics, for it is well known that global symmetries are never exact.

The presence of anomalies, as in the case of T-parity \cite{Hill:2007zv}, or R-parity violating terms in supersymmetry \cite{Barbier:2004ez} would often mean that the dark matter candidates arising from these theories are neither stable nor long-lived in the cosmological sense.  Even if this had not been the case, the presence of gravity would necessarily induce the violation of global symmetries as was first revealed in studies of black holes \cite{b1, b2, b3, b4}. So the lightest particle charged under a particular global symmetry would have, at best, a very long lifetime. Indeed, it has even been conjectured that discrete global symmetries are violated maximally by gravity \cite{Banks:1989ag, Krauss:1988zc}.

Additional motivation can be found in numerical simulations of the universe (based on the conventional $\Lambda$CDM cosmology) which predict an overabundance of substructures as compared to actual observations. Models with decaying dark matter \cite{Cen:2000xv, Borzumati:2008zz} provide an extremely compelling and natural mechanism for suppressing the power spectrum at small scales thus resolving the discrepancy.

Continuing on the line of thought leading from particle physics to cosmology, the question that then naturally springs to mind is ``What can cosmology say about decaying dark matter and the particle physics theories that contain them?" It is this intriguing prospect that we explore in this paper.

There have been a few papers \cite{Ichiki:2004vi, Gong:2008gi, Lattanzi:2008zz} in recent years analyzing DM decay into electromagnetically non-interacting particles using just the cosmic microwave background data (Ref.\cite{Gong:2008gi} also includes supernova data). In this paper, we revisit this scenario using a Markov Chain Monte Carlo (MCMC) analysis employing all available datasets from the cosmic microwave background (CMB), Type Ia supernova (SN), Lyman-$\alpha$ forest (Ly$\alpha$), large scale structure (LSS) and weak lensing (WL) observations. We find that the lifetime of decaying DM is constrained predominantly by the late time Integrated Sachs Wolfe (ISW) effect to be $\Gamma^{-1} \gtrsim 100$ Gyr. In the main body of the paper, we will comment on the discrepancies between the results of Refs.\cite{Ichiki:2004vi, Gong:2008gi, Lattanzi:2008zz}.

The studies in the preceding paragraph considered only the case where there was negligible reionization of the universe due to DM decay. In an attempt to address this, Ref.\cite{Zhang:2007zzh} analyzed the scenario of DM decaying into only electromagnetically interacting products, that get partially absorbed by the baryonic gas, using a subset of the available CMB datasets. Our paper extends their analysis by using all the available CMB datasets, and also the SN, Ly$\alpha$, LSS and WL datasets. Besides the smaller selection of datasets, their analysis also ignores the impact of DM decay on cosmological perturbations which renders it ineffectual in the parameter space where there is negligible reionization. Our treatment allows the decay products to not only reionize the universe but also takes into account the effect of DM decay on cosmological perturbation. This allows us to generate many other observables, particularly, the late time ISW effect that is crucial to constrain the lifetimes at low reionization. Another key difference between the analyses is that we use a combined reionization parameter for both DM reionization and phenomenological star formation reionization, rather than just treating them separately as was done in Ref.\cite{Zhang:2007zzh}, because current observations cannot distinguish which contribution to reionization is the dominant one. Doing the MCMC analysis, we find that the lifetime of decaying DM in this scenario constrained to be $ (f \Gamma) ^{-1} \gtrsim 5.3 \times 10^8$ Gyr, where $f$ is a phenomenological parameter introduced by Ref.\cite{Zhang:2007zzh} and related to the degree of reionization.

Astrophysical constraints together with additional assumptions  have also been used \cite{PalomaresRuiz:2007ry, Yuksel:2007dr}  to give even tighter bounds than ours on the lifetime of the decaying DM. While interesting and complementary, these lie outside the scope of our present paper.

Having obtained the bounds on the lifetime of decaying dark matter, we will then explore the implications of our cosmological analysis on particle physics models beyond the Standard Model. We will present a complete list of cross-sections for spin-0, spin-1/2 and spin-1 dark matter to decay into Standard Model degrees of freedom via effective operators. Obviously, this can be easily extended to other models with additional light degrees of freedom (for instance, hidden valley models \cite{Strassler:2006im}) by appropriate substitution of the parameters. Applying the bound on the lifetime of the decaying DM, we can then place limits on the size of the parameters of theories. For generic theories with a decaying dark matter of $\sim 100$ GeV mass, the coupling constant in the effective dimension-4 operators responsible for dark matter decay will be shown to be $\lesssim 10^{-22}$. We will also look at specific representative cases of theories beyond the Standard Model physics and investigate the possibility of viable dark matter candidates: the spin-0 messenger DM in the context of gauge mediation messenger number violation, the spin-1/2 bino DM in the scenario with R-parity violation and the spin-1 ``massive photon partner''DM in the framework of T-parity violation.



The rest of this paper is organized as follows. In Section 2, we discuss the physics of decaying dark matter cosmology as well as introduce the datasets that we will be using. Section 3 contains our Markov Chain Monte Carlo results and discussions of the cosmological implications. In Section 4, we explore the consequences of these results for particle physics theories by enumerating the decay channels and partial widths. Representative models from theories of gauge-mediated supersymmetry breaking, minimal supergravity and little Higgs were also investigated using the results of our analysis. We conclude and briefly comment on future prospects in Section 5.

\section{Decaying Cold Dark Matter Cosmology}

We will assume the standard picture of $\Lambda$CDM cosmology, i.e. a Friedman-Robertson-Walker universe that is principally composed of dark energy and cold dark matter, with one crucial modification; that is, we have a cold dark matter that is very long-lived but ultimately decays. As we are considering  lifetimes of gigayears (Gyr), the fraction of DM decays happening during or soon after big bang nucleosynthesis (BBN) is negligible and hence would not alter the predictions of BBN. To perform a model-independent analysis, we allowed decays to all possible SM particles. However, we will assume that the long term decay products are relativistic. While we include branching ratios to intermediate non-relativistic states, they are assumed to be short-lived and will rapidly decay into light relativistic degrees of freedom.

The evolution of background and first order perturbation in decaying cold dark matter model was first formulated in longitudinal gauge \cite{Ichiki:2004vi}, which means the decay rate has to be treated with care as the CDM is not at rest in the longitudinal gauge. We, on the other hand, will work in CDM rest frame using synchronous gauge with the line element written as
\begin{equation}
ds^2=a^2(\tau)[-d\tau^2+(\delta_{ij}+h_{ij})dx^idx^j] .
\end{equation}
where $\tau$ is conformal time, and $t$ the cosmological time ($dt=a(\tau)d\tau$). In this paper we follow the convention $a=1$ today.

The decay equation
\begin{equation} \label{eq_decay}
\frac{d\rho_{cdm}}{dt}=-\Gamma\rho_{cdm},
\end{equation}
can be  reformulated in a covariant form
\begin{equation} \label{eq_decaycov}
T^\mu_{\ \nu;\mu}(\texttt{CDM})=G_\nu,
\end{equation}
where the force density vector $G_\nu$ can be calculated from its value in CDM rest frame
\begin{equation}
G_\nu|_{\texttt{CDM rest}} = (-\Gamma \rho_{cdm}, 0, 0, 0).
\end{equation}

The conservation of total energy momentum tensor requires
\begin{equation}
T^\mu_{\ \nu;\mu}(\texttt{dr})=-G_\nu,
\end{equation}
where the daughter radiation (dr) is composed of the CDM decay products.

The equations describing the evolution of background are
\begin{eqnarray}
\dot\rho_{cdm} &=& -3{\cal H}\rho_{cdm}-a\Gamma \rho_{cdm}, \\
\dot\rho_{dr} &=& -4{\cal H}\rho_{dr}+a\Gamma\rho_{cdm},
\end{eqnarray}
where dot denotes the derivative with respect to conformal time $\tau$.  We have defined the conformal expansion rate to be ${\cal H} \equiv \frac{\dot a}{a}$.

We will only consider scalar metric perturbations, which in Fourier space can be expanded as following \cite{Ma:1995ey}.
\begin{equation}
h_{ij}({\mathbf x},\tau)=\int d^3{\mathbf k}e^{i{\mathbf k}\cdot{\mathbf x}}[{\mathbf n_i}{\mathbf n_j}h({\mathbf k},\tau)+6({\mathbf n_i}{\mathbf n_j}-\sfrac{1}{3}\delta_{ij})\eta({\mathbf k},\tau)],
\end{equation}
where ${\mathbf n}\equiv {\mathbf k}/|{\mathbf k}|$.

Our choice of gauge and coordinates lead to the following simple density perturbation equation for CDM,
\begin{equation}
\dot\delta_{cdm}=-\frac{1}{2}\dot h.
\end{equation}
The terms containing $\Gamma$ all cancel out, because the background density and overdensity are decaying with the same rate.

Instead of using simple hydrodynamic approximation for the decay product \cite{Kofman:1986am}, which might give correct order of magnitude but less accurate results,
we use the full Boltzmann equations to describe the decay product, which were first given by Ref.\cite{Bond1984}, and recently updated by \cite{Lattanzi:2008zz} for decaying DM cosmology,
\begin{eqnarray}
\dot\delta_{dr} &=& -\sfrac{2}{3}\dot h-\sfrac{4}{3}kv_{dr}+ a \Gamma \frac{\rho_{cdm}}{\rho_{dr}}(\delta_{cdm}-\delta_{dr}), \\
\dot v_{dr} &=& k(\sfrac{1}{4}\delta_{dr}-\sfrac{1}{2}\Pi_{dr})-a\Gamma\sfrac{\rho_{cdm}}{\rho_{dr}}v_{dr}, \\
\dot\Pi_{dr} &=& k(\sfrac{8}{15}v_{dr}-\sfrac{3}{5}F_3)+\sfrac{4}{15}\dot h +\sfrac{8}{5}\dot\eta -a\Gamma\frac{\rho_{cdm}}{\rho_{dr}}\Pi_{dr}, \\
\dot F_l &=& \frac{k}{2l+1}[lF_{l-1}-(l+1)F_{l+1}]-a\Gamma\frac{\rho_{cdm}}{\rho_{dr}}F_l,
\end{eqnarray}
where $l=3,4,5,...,$, $F_2=\Pi$ and for the rest, we have used the conventions of Ref.\cite{Ma:1995ey}. Because CDM particles are heavy and non-relativistic, we have treated the CDM as a perfect fluid.

For the case where the DM candidate also decays into electromagnetically interacting particles (e.g. photons or electron/positrons), we have to be more careful. This is because the decays may deposit significant energy into baryonic gas and contribute to the reionization of universe. Following \cite{Chen:2003gz, Zhang:2007zzh, Schleicher:2008dt}, we introduced a phenomenological factor $f$ as the fraction of the decay energy deposited in the baryonic gas. For long-lifetime dark matter models, the reionization due to dark matter decay only depends on the combination $\zeta=f \, \Gamma/H_0$. We use $\zeta$ as an additional parameter in our MCMC analysis. Without a prior on $f$, the constraint on $\zeta$ does not directly give any information on $\Gamma$. However, for given dark matter models, one should in principle be able to calculate the decay branches, and therefore give a rough estimate for $f$ under certain additional assumptions. Following Ref.\cite{Zhang:2007zzh, Schleicher:2008dt}, we modify RECFAST \cite{Seager1999, Seager2000} to calculate the reionization due to DM decay. In this scenario, the reionization is dominated by DM decay at redshift $z>20$, and is competing with the contribution from star formation (or other sources) at some redshift between $z=6$ and $z=20$. Without knowing the details of star formation or other reionization sources, we use the following phenomenological model, which can be regarded as a combination of CosmoMC phenomenological formula and DM decay reionization formula,

\begin{equation}
x_e=\max \{ x_e\textsuperscript{RECFAST}, \frac{1+f_{He}}{2}[1+\tanh(\frac{(1+z)^{1.5}-(1+z_{re})^{1.5}}{1.5\sqrt{\Delta z} })]\},
\end{equation}
where $x_e$, the ionized fraction, is defined as the ratio of free electron number density to hydrogen number density; $f_{He}$ is the ratio of helium number density to hydrogen number density; $x_e$\textsuperscript{RECFAST} is the modified RECFAST output ionized fraction (i.e., the ionized fraction assuming DM decay is the only source of reionization; $\Delta z$ is the redshift width of reionization (due to other sources), for which we have taken the CosmoMC \cite{Lewis2002} default value $0.5$; the last free parameter, reionization redshift $z_{re}$, is determined by the total optical depth $\tau_{re}$.

With all the above equations, we modified CosmoMC to analyze the decaying CDM model. In addition, we also incorporated weak lensing data into the Markov Chain Monte Carlo (MCMC) analysis. The datasets used in this paper are listed below. For each dataset, we either wrote a new module to calculate the likelihood or modified the default CosmoMC likelihood codes to include the features of the decaying CDM model.

\begin{description}

\item{\it Cosmic Microwave Background (CMB)}

We employ the CMB datasets from WMAP-5yr \cite{Hinshaw2008,Komatsu2008},
BOOMERANG \cite{Jones2006,Piacentini2006,Montroy2006}, ACBAR
\cite{Reichardt2008,Kuo2006,Runyan2003, Goldstein2003}, CBI
\cite{Pearson2003,Readhead2004a,Readhead2004b,Sievers2007}, VSA
\cite{Dickinson2004}, DASI \cite{Halverson2002,Leitch2005}, and
MAXIMA \cite{Hanany2000}. Also included are the Sunyaev-Zeldovich (SZ) effect for WMAP-5yr, ACBAR
and CBI datasets. The SZ template is obtained from hydrodynamical simulation \cite{Bond2005}.
When calculating the theoretical CMB power spectrum, we have also turned on CMB lensing in CosmoMC.

\item{\it Type Ia Supernova (SN)}

The Union Supernova Ia data (307 SN Ia samples) from The
Supernova Cosmology Project \cite{Kowalski2008} was utilized. For
parameter estimation, systematic errors were always included.

\item{\it Large Scale Structure (LSS)}

For large scale structure we will use the combination of
2dFGRS dataset \cite{Cole2005} and SDSS
Luminous Red Galaxy Samples from SDSS data release 4 \cite{Tegmark2006}.
It should be noted that the power spectrum likelihood already contains the
information about BAO (Baryon Acoustic Oscillation \cite{Eisenstein2005,Percival2007}).

\item{\it Weak Lensing (WL)}

Five weak lensing datasets were employed in
this paper. The effective survey area and galaxy number density of
each survey are listed in Table~\ref{tblwldata}.
\begin{table}[h]
  \centering
  \caption{Weak Lensing Data}\label{tblwldata}
  \begin{tabular}{|c|c|c|}
  \hline
  Data & $A_{eff}($deg$^{-2}$) &
$n_{eff}($arcmin$^{-2}$) \\
    \hline
COSMOS \cite{Massey2007,Lesgourgues2007} & 1.6 & 40\\
\hline
CFHTLS-wide \cite{Hoekstra2006,Benjamin2007,Schimd2007}& 22 & 12 \\
\hline
GaBODS \cite{Hetterscheidt2006,Hetterscheidt2007,Benjamin2007}& 13 & 12.5 \\
\hline
RCS \cite{Hoekstra2002a,Hoekstra2002b,Benjamin2007}& 53 & 8 \\
\hline
VIRMOS-DESCART \cite{Van-Waerbeke2005,Schimd2007,Benjamin2007}& 8.5 &  15 \\
\hline
  \end{tabular}
\end{table}

For COSMOS data we used the CosmoMC module written by Julien
Lesgourgues\cite{Lesgourgues2007} . For the other four weak lensing datasets, we utilized the likelihood
given in Benjamin \etal \cite{Benjamin2007}. We take the best fit
parameters for the following galaxy number density formula,
\begin{equation}
n(z)=\frac{\beta}{z_0\Gamma(\frac{1+\alpha}{\beta})}(\frac{z}{z_0})^\alpha
\exp{(-\frac{z}{z_0})^\beta}.
\end {equation}

Benjamin \etal give two sets of best fit parameters, fitting on
the galaxy samples with median photometric redshift $0<z_p<4$ and
$0.2<z_p<1.5$, respectively (see Fig. 2 and Table 2 in
\cite{Benjamin2007}). We only used the data from the $0<z_p<4$ region. We also
simplified the marginalization on $n(z)$ parameters by assuming a Gaussian
prior on $z_0$. The width of Gaussian prior is adjusted so that
the mean redshift $z_m$ has an uncertainty of $0.03 (1+z_m)$, i.e.
\begin{equation}
\sigma_{z_0}=0.03 \big
[\frac{\Gamma(\frac{1+\alpha}{\beta})}{\Gamma(\frac{2+\alpha}{\beta})}+
z_0\big ].
\end{equation}

The weak lensing data only measure matter power spectrum at angular scales less than a few degrees, which corresponds to scales less than a few hundred Mpc. This is much less than the Jean's length of the daughter radiation and therefore we can ignore the daughter radiation when calculating the power spectrum of projected density field.
\begin{eqnarray}
P_l(\kappa) &=&(\frac{4\pi G}{c^4})^2\int_0^{\chi_H}\rho_m^2a^4P_{3D}(\frac{l}{d_{cA}(\chi)};\chi) \nonumber \\
&& \times \big[\int_\chi^{\chi_H}d\chi' n(\chi') \frac{d_{cA}(\chi'-\chi)}{d_{cA}(\chi')}\big]^2.
\end{eqnarray}
 We should stress that this is specific to the decaying CDM model and differs from the conventional CDM model \cite{Kaiser1992, Kaiser1998}.

\item{\it Lyman-$\alpha$ Forest}

The following Ly$\alpha$ forest datasets were applied.

\begin{enumerate}
    \item {The dataset from Viel \etal \cite{Viel2004} consist of LUQAS
sample \cite{Kim2004} and the Croft \etal data \cite{Croft2002}. }

    \item {The SDSS Ly$\alpha$ data presented in McDonald \etal
\cite{McDonald2005,McDonald2006}. To calculate the likelihood, we
interpolated the $\chi^2$ table in the three-dimensional amplitude-index-running space.}
\end{enumerate}

\end{description}

We explore the likelihood in nine-dimensional parameter space, i.e., the Hubble parameter $h$, the baryon density $\Omega_b h^2$, the amplitude and index of primordial power spectrum ( $A_s$ and $n_s$), the DM decay reionization parameter $\zeta$, the total reionization optical depth $\tau_{re}$, the SZ amplitude $A_{SZ}$, the decay rate normalized by Hubble parameter $\frac{\Gamma}{H_0}$, and the CDM density in early universe $\Omega_{cdm, e} h^2$. The parameter $\Omega_{cdm,e}$ is defined to be

\begin{equation}
\Omega_{cdm,e}\equiv \frac{(\rho_{cdm} a^3)|_{a \ll 1}}{\rho_{crit0}},
\end{equation}
where $\rho_{crit0}\equiv \frac{3H^2}{8\pi G}$ is today's critical density. As the CDM in our case decays, we made a distinction between $\Omega_{cdm,e}$ and  $\Omega_{cdm}$ where the latter is defined to be the usual fractional CDM density today  ($\rho_{cdm0}/\rho_{crit0}$).

\section{Markov Chain Monte Carlo Results and Discussion}

\begin{figure}
\includegraphics{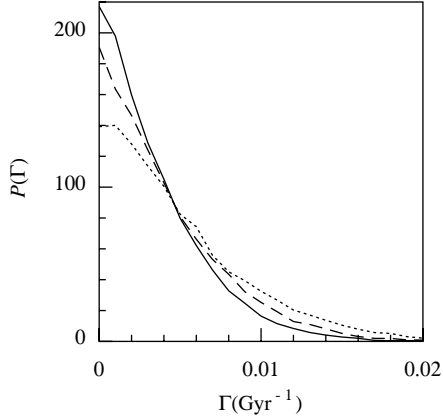}
\caption{Posterior probability density function of the decay rate $\Gamma$. Solid line: using all the datasets. Dashed line: CMB + SN + LSS + Ly$\alpha$. Dotted line: CMB only. The probability density function is normalized as $\int P(\Gamma)d\Gamma =1$.}  \label{fig1d}
\end{figure}

For the case with negligible reionization, we generated 8 MCMC chains, each of which contains about 3000 samples. The posterior probability density function of CDM decay rate can be directly calculated from the Markov Chains, as shown in Fig. \ref{fig1d}. The corresponding 68.3\% and 95.4\% confidence level lower bounds on lifetime are $\Gamma^{-1} \gtrsim 230$Gyr and $\Gamma^{-1} \gtrsim100$Gyr, respectively. If we take the lifetime of universe to be 14Gyr, the 95.4\% confidence level limit (i.e. lifetime 100Gyr) corresponds to a scenario that roughly 15\% of CDM has decayed into radiation by today.

\begin{figure}
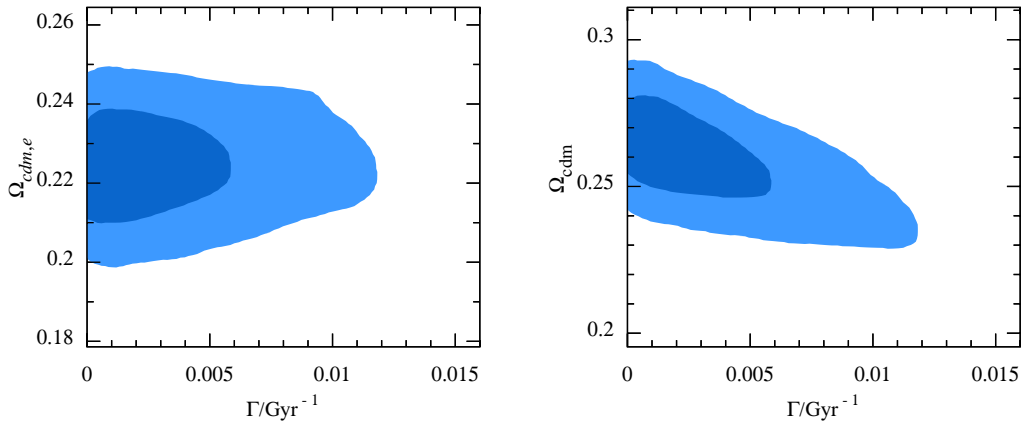

\includegraphics{DDM_early_2D.eps} %
\includegraphics{DDM_2D.eps}
\caption{Constraints on the early universe CDM density parameter $\Omega_{cdm,e}$ and decay rate $\Gamma$, using all the datasets, is plotted on the left panel. For comparison, the present day CDM density parameter $\Omega_{cdm}$ and decay rate $\Gamma$ is plotted on the right panel. The inner and outer contours correspond to 68.3\% and 95.4\% confidence levels, respectively.} \label{fig2d}
\end{figure}

In our analysis, all the early universe physics before recombination remains unchanged. The CMB power spectrum is however significantly modified due to two effects. One is that the decay of CDM modifies the evolution of background, which results in a different distance to last scattering surface compared to the conventional case. The second one is that the decay of CDM affects the cosmological perturbations in late universe, resulting in an enhancement of the integrated Sachs-Wolfe (ISW) effect beyond that due to the cosmological  constant. And it is this effect, anticipated by Kofman \etal\cite{Kofman:1986am}, that gives us the most restrictive bound on the lifetime of decaying dark matter for the scenario with negligible reionization.

\begin{figure}
\includegraphics[width=3.5in] {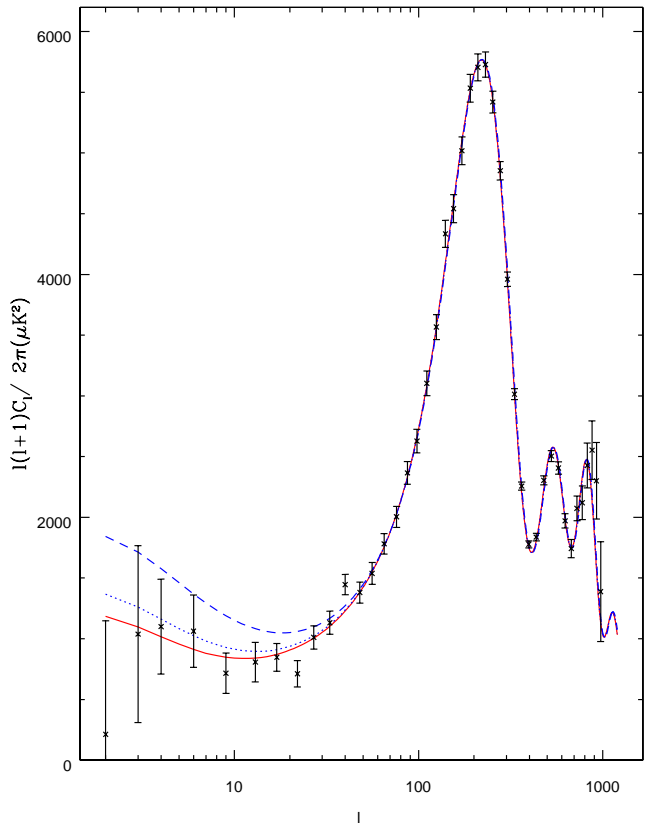}
\caption{CMB power spectrum for different dark matter decay rate, assuming the decayed particles are relativistic and weakly interacting. For the CDM density parameter, we choose $\Omega_{cdm,e}h^2$ to be the same as WMAP-5yr median $\Omega_mh^2$. For the other cosmological parameters we use WMAP-5yr median values. By doing this, we have fixed the CDM to baryon ratio at recombination. In a similar plot in Ichiki \etal \cite{Ichiki:2004vi} $\Omega_{cdm0}h^2$ is instead fixed. Therefore the height of first peak, which has dependence on CDM to baryon ratio at recombination, will significantly change as one varies the decay rate. In this plot the red line corresponds to a stable dark matter . The blue dotted line corresponds to dark matter with a lifetime 100 Gyr, and the blue dashed line 27 Gyr. The data points are WMAP-5yr $<T  T>$ spectrum mean values and errors (including instrumental errors and cosmic variance).} \label{figcls}
\end{figure}

Let us review the inconsistency between past papers on this issue. We start with Ref.\cite{Gong:2008gi}. Now, CMB and SN observations today can measure the fractional CDM density to a roughly 15\% level \cite{Komatsu2008} (within 95\% confidence level). We will expect the constraints on CDM decay ratio to be the same order of magnitude. This simple estimation does not take into account the fact that the decayed product still forms part of the matter component (the equation of state is changed to $\frac{1}{3}$), and that the DM decay happens mostly at low redshift.  Therefore if we do not take the cosmological perturbation into account, the data should allow about 15\% dark matter to have decayed by today, i.e., we should not get a bound better than $100$ Gyr. This simple analysis implies that the recent lower bound of lifetime ($\Gamma^{-1}>700$Gyr at 95.4\% confidence level) obtained by Ref.\cite{Gong:2008gi}, which does not take into consideration the impact of DM decay on cosmological perturbation (the location of first CMB peak is affected only through the change of background evolution), may not be credible. If indeed the CDM lifetime is 700Gyr, only about 2\% of CDM has decayed into radiation by today, and by the time of recombination, less than $10^{-6}$ of CDM has decayed. The change in background evolution is so tiny that it should not be detectable by current cosmological data.

To compare with Ichiki \etal \cite{Ichiki:2004vi}, we re-did the analysis using just the CMB datasets, and found the CDM lifetime $\Gamma^{-1}\gtrsim70$ Gyr at 95.4\% CL, which is consistent with their results. The reason we obtained a more constrained value than their $\Gamma^{-1}\gtrsim 52$ Gyr at 95.4\% C.L. is probably because we used WMAP-5yr compared with their WMAP-1yr dataset. We expect the WMAP-9yr dataset, when published and analyzed, to exhibit only a modest improvement because the information from the late time ISW effect is limited by cosmic variance. Recently, Lattanzi \etal \cite{Lattanzi:2008zz} obtained a bound of $\Gamma^{-1} \gtrsim 250$ Gyr at 95.4\% C.L. with just the WMAP-3yr data, which is not consistent with both Ichiki \etal and our results. We notice that in the Fig. 3 of their paper, the proximity between 68.3\% and 95.4\% confidence level bounds on $\Gamma$ indicates a sudden drop of marginalized likelihood ${\cal L}(\Gamma)$. In our result, as shown in Fig. \ref{fig1d}, this sudden drop feature is not seen.

\begin{figure}
\includegraphics{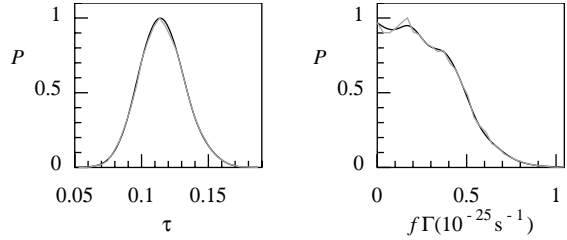}%
\caption{The marginalized posterior likelihood of the total optical depth and that of the DM decay reionization parameter.} \label{reion}
\end{figure}

\begin{figure}
\includegraphics{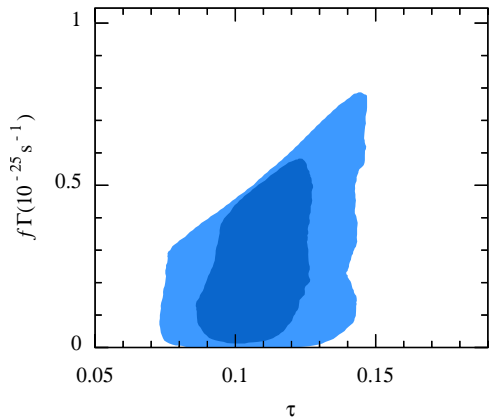}
\caption{The marginalized 2D likelihood contours. The inner and outer contours correspond to 68.3\% and 95.4\% confidence levels, respectively.} \label{reionb}
\end{figure}

Let us move on to the scenario where there is significant reionization due to the decaying dark matter. We generated another 8 MCMC chains, each of which contains about 6000 samples.
The results of our analysis can be seen in Fig. \ref{reion} and \ref{reionb}, where we show the constraints on DM decay reionization parameter. A few things should be noted. Firstly, the sharp boundary (reflected in the closeness of the two contours) on the rising edge in Fig. \ref{reionb} is due to the fact that for a given $\tau_{re}$, DM decay has an upper limit because the optical depth due to DM decay should not extend beyond $\tau_{re}$. Secondly, the plateau of likelihood around $f \, \Gamma=0$ in Fig. \ref{reion} indicates that current CMB polarization data can only constrain the total optical depth, but cannot distinguish between DM decay reionization and star formation reionization. In other words, the data does not favor or disfavor DM decay reionization, as long as its contribution to total optical depth is not larger than the preferred $\tau_{re}$.

The constraint we have obtained is $f \, \Gamma \lesssim 0.59\times 10^{-25}$s$^{-1}$ at 95.4\% confidence level. This result is about a factor of 3 better than Zhang \etal \cite{Zhang:2007zzh}. In the limit where reionization is negligible, Zhang \etal cannot give a strong bound on $\Gamma$ because they have ignored the impact of DM decay on cosmological perturbations. Hence, their constraint on DM decay is essentially, only from CMB polarization data. Our analysis, which combines many different cosmological datasets and includes the calculation of the impact of DM decay on all the observables, gives a stringent constraint on $\Gamma$ even in the $f=0$ limit. As for the limit of significant reionization, our bounds are, as mentioned earlier, an improvement over Zhang \etal and this may be due to the fact that we have used more datasets. However, the priors of the parameters may also alter the result. A notable difference between the models is due to the fact that one of the parameters they have adopted, the optical depth without DM decay, is ill-defined in our model. Furthermore in their model, the cutoff of DM decay reionization at $z=7$ was explicitly chosen. These differences might have led to Fig. 1 in their paper which shows a preference for a zero dark matter decay rate, a feature that is absent in our results.

The constraint we have obtained is $f \, \Gamma \lesssim 0.59\times 10^{-25}$s$^{-1}$ at 95.4\% confidence level. This result is about a factor of 3 better than Ref.\cite{Zhang:2007zzh}. In the limit where reionization is negligible, Ref.\cite{Zhang:2007zzh} cannot give a strong bound on $\Gamma$ because they have ignored the impact of DM decay on cosmological perturbations. Hence, their constraint on DM decay is essentially only from CMB polarization data. Our analysis, which combines many different cosmological datasets and includes the calculation of the impact of DM decay on all the observables, gives a stringent constraint on $\Gamma$ even in the $f=0$ limit. As for the limit of significant reionization, our bounds are, as mentioned earlier, a significant improvement over Ref.\cite{Zhang:2007zzh}. The difference may be due to the fact that we have used more datasets. However, the priors of the parameters may also alter the result. A notable difference between the models is due to the fact that one of the parameters they have adopted, the optical depth without DM decay, is ill-defined in our model. Also in their model, the cutoff of DM decay reionization at $z=7$ was explicitly chosen. These differences might have led to Fig. 1 in their paper which shows a preference for a zero dark matter decay rate, which is absent in our analysis.

\section{Implications for Particle Physics Models with Decaying Cold Dark Matter}

Our results most certainly impose constraints on extensions of the Standard Model of particle physics (SM) with DM candidates. Making the assumption that DM decays into SM fields, we will investigate all the probable decay channels unless forbidden by either symmetry or kinematics, or highly suppressed by phase space considerations. We then sum up their partial decay widths to obtain the functional form for the lifetime of each of the decaying DM candidates. One might be worried about including even the decays to non-relativistic particles as that might invalidate our earlier assumption that cosmologically, the dark matter decay products are relativistic. While it is true that the DM particle can and will decay (provided it is not kinematically forbidden) into non-relativistic massive gauge bosons or heavy quarks, these heavy particles themselves will be assumed to subsequently decay very rapidly into much lighter particles of the SM that will be relativistic. Obviously in specific models, certain channels could be expressly forbidden by symmetries and this can also be handled by our analysis.

To obtain the functional form for the lifetime of the decaying DM, we will approach it from the point of view of effective field theory. We will do a model independent analysis by considering generic Lagrangian terms for these decays with the corresponding coupling constants acting as Wilson coefficients. Since the DM is electrically neutral, the total charge of decay products should also be zero. Moreover, the decay rates for different channels are dependent on the intrinsic spin of the DM because of possible spin-dependent couplings. Below we discuss the decays of DM with spin 0, 1/2 and 1. We will only consider decay processes of the lowest order, as higher-order processes involve more vertex insertions and so are assumed to be suppressed. Additionally in our approach, we will work in the framework where all the gauge symmetries (including those of Grand Unified Theories if present) except for $SU(3)_{c} \times U(1)_{em}$ are broken and the effects encoded in the coupling constants of the effective operators. This can potentially give rise to naturally very small coupling constants as they could contain loop factors or powers of very small dimensionless ratios. This is a more cost-effective and model-independent way of taking into account the myriad possibilities of UV-completing the Standard Model of particle physics.

Having obtained the functional form of the lifetime in terms of the fundamental parameters of the underlying particle physics models, we can then compare it it with the numerical value obtained from the cosmological analysis of the previous section. This would allow us to place definitive bounds on the fundamental parameters of candidate models for the particle physics theory beyond the Standard Model. It should be noted that we will be using the most conservative 95.4\% confidence level bound on the lifetime of the decaying dark matter, i.e. without significant reionization. To assume otherwise would require a more complete knowledge of the ionization history of the universe than is currently understood.

Let us now proceed to the case of a generic scalar DM candidate and see how the above ideas are implemented.

\subsection{Spin-0 Dark Matter}

We first consider a spin-zero DM candidate, $S$. Decays into right-handed neutrinos and left-handed anti-neutrinos are not considered as the former may be more massive than $S$. Even if they are light enough for $S$ to decay into, we expect the decay into SM (anti-) neutrinos to be dominant. For this decay channel, we can proceed in the same way as in the case of neutral pion decay. In the SM, (anti-) neutrinos couple to other matter in the form of the chiral current $\bar{\nu_L}\gamma^\mu\nu_L$ since $S$ carries no Lorentz index. The lowest dimension operator responsible for this decay will be of the form
 $g_0\bar{f}\gamma^\mu(1+r_a\gamma_5)f\partial_\mu S/\Lambda$, where we parameterize $g_0$ as the coupling constant of dimension zero, and $\Lambda$ is some cutoff scale. The presence or absence of $\gamma_5$ in the operator depends on whether $S$ couples to the SM neutrinos in a vectorial or axial-vectorial way. The corresponding decay rate is given by

\begin{equation}\label{piondecayrate_will}
\Gamma = \frac{g_0^2r_a^2}{2\pi}\frac{m_f^2m_S}{\Lambda^2}\sqrt{1-4\frac{m_f^2}{m_S^2}},
\end{equation}
where $m_f$ is the mass of the decay product. Here we have assumed that (anti-) neutrinos have Dirac mass.

We focus our attention on DM with mass $m_S\gg m_f$. Then for decay products such as (anti-) neutrinos (or other light SM particles), it is safe to make the approximation $1 - 4m_f^2/m_S^2\simeq 1$. Since the neutrino is left handed, we take $r_a=-1$ . If $S$ decays dominantly into $\nu_e\bar{\nu_e}$, our lower bound on $\Gamma$ then constrains the following parameter,
\begin{eqnarray}\label{constraint1}
\frac{g_0^2 m_f^2 m_S}{\Lambda^2} &\lesssim&  1.3 \times 10^{-42} \mathrm{GeV} \ (95.4 \% \textrm{ confidence level}),
\end{eqnarray}
where we have used $1 \mathrm{Gyr^{-1}} = 2.087 \times 10^{-41} \mathrm{GeV}$.

Here we can see that helicity suppression at work. When the mass of the decaying particles is very small, the decay of the spin-zero DM candidate will be suppressed as expected. The presence of helicity suppression gives us a value of $g_0$ that is larger than in most other cases, as we will see. For example, if the mass of the DM candidate is $m_S \sim 100 \mathrm{GeV}$, the neutrino mass around $m_f \sim 2$eV, and the cutoff is $\Lambda \sim 10 \mathrm{TeV}$, then the coupling constant $g_0$ has to be $\sim 10^{-11}$. If on the other hand the coupling constant $g_0$ is $\mathcal{O}(1)$ and we take the same values of the neutrino masses and of the DM candidate, then we have that $\Lambda \sim 10^{13} \rm GeV$.

Table \ref{table:spin0} in the Appendix lists out possible Lagrangian terms for the decay of $S$ into SM particles, and the corresponding decay rates, summing over final state spins. Apart from focusing solely on the $S\to\nu\bar{\nu}$ channel, we can consider scenarios in which all the interaction terms in Table \ref{table:spin0} are present to contribute to the decay rate of $S$, with all coupling constants of the same order of magnitude, $g_0$. For simplicity, we will also assume $m_{decay \ product}/m_S$ is negligible compared to 1. This would immediately imply that the helicity-suppressed term, $g_0\bar{f}\gamma^\mu(1+r_a\gamma_5)f\partial_\mu S/\Lambda$, gives rise to insignificant decay rate when compared to other terms. So the most relevant terms are the ones that come from the operator $g_0 S\bar{f}(1+ir_p\gamma_5)f$. Then we have
\begin{eqnarray}
\sum_{f\in\ {\rm SM}}\Gamma(S\to f\bar f) &\approx& \frac{g_0^2m_S}{8\pi}(1+r_p^2)\sum_{f\in\ {\rm SM}}N_f\nonumber \\
&=& \frac{21g_0^2m_S}{8\pi}(1+r_p^2).
\end{eqnarray}
Here the decay to $t\bar t$ is not included as this channel may not be kinematically feasible for a DM particle of $\sim 100 \mathrm{GeV}$. The parameter $r_p$ is also assumed to be the same for all $f$. In a similar way, rates for the other decays into SM gauge bosons can be worked out:
\begin{eqnarray}
& &\Gamma(S \to \gamma\gamma)+\Gamma(S\to gg)+\Gamma(S\to ZZ)+\Gamma(S\to W^+W^-)+\Gamma(S\to Z\gamma)\nonumber \\
&\approx &\frac{g_0^2m_S}{64\pi}\left (80+640+82+81+10\right ).
\end{eqnarray}
Our result imposes an upper bound on the sum of decay rates via all the channels. Taking $r_p=0$ as an example, our result would give us the constraint
\begin{equation}
g_0^2m_S\lesssim 3.9\times 10^{-44} \mathrm{GeV} \ (95.4 \% \textrm{ confidence level}).
\end{equation}

\subsubsection{Messenger number violation in gauge-mediated supersymmetry breaking theories}

Let us now investigate the messenger parity in gauge-mediated supersymmetry breaking theories. In these models, there could potentially be a dark matter candidate coming from the electromagnetically-neutral scalar field that is formed from the $SU(2)$ doublets of the $5$ and $\bar{5}$ of the messenger sector \cite{Dimopoulos:1996gy}. However, it is usually not easy to realize this because the lightest odd-messenger parity particle (LOMPP) often turns out not to be the electromagnetically-neutral field that we require. It has been claimed in the same paper that certain F-terms would lift the degeneracy. If we further assume that it does not significantly modify the effective low energy theory, the analysis becomes very  much model-independent as there are only certain couplings that lead to decay of the LOMPP.
Following \cite{Dimopoulos:1996gy}, the Kahler potential  is given by
\begin{equation}
K=\int d^4 \theta \left( \overline{5}^\dag_M\overline{5}_M +5_M^\dag5_M+\overline{5}^\dag_F\overline{5}_F +10_F^\dag10_F\right)+
\frac{g_0}{Mp}\left(\overline{5}^\dag_M10_F^2 +5_M^\dag\overline{5}_F10_F+h.c.\right),
\end{equation}

\noindent where $\overline{5}_M$ and $5_M$ are the messengers and $\overline{5}_F$ and $10_F$ are the ordinary superfields. The terms that are Planck suppressed are the ones that violate messenger number by one unit.

As for the superpotential, we have
\begin{equation}
W=\int d^2\theta \, \rho \, \overline{5}_M5_M+\frac{g_0'}{Mp}\overline{5}_M10^3_F,
\end{equation}

\noindent where $\rho$ is the supersymmetry breaking spurion field and once again the terms that are Planck suppressed are dimension-5 messenger number violating terms. Without full knowledge of the UV-sensitive physics (F-terms that lift the other fields while retaining a viable LOMPP), we can still give an estimate of the order of magnitude of the decay rate of the LOMPP,
\begin{equation}
\Gamma \sim N\frac{g_0^2m_{mess}^3}{M_p^2\pi}F_k,
\end{equation}

\noindent where $N$ are the different degrees of freedom that the LOMPP can decay into. $F_k$ is a function that contains the kinematic information and we will assume that is close to one.
We can then put a constraint on the coupling constant and on the messenger mass. Since the lifetime is $100 \mathrm{Gyr}$, then for $F_k \sim 1$ and $N\sim100$, we have
\begin{equation}
g_0^{2}\left(\frac{m_{mess}^3}{M_p^2} \right)\lesssim 6\times 10^{-45} \mathrm{GeV},
\end{equation}

\noindent where we have assumed one universal coupling constant $g_0$. For the sake of discussion, if we consider a coupling constant of order one, we get a small messenger mass $m_{mess} \sim 0.02 \mathrm{GeV}$. This can be improved if we go to dimension six operators which gives a generic decay rate of the form

\begin{equation}
\Gamma \sim \frac{Ng_0^2m_{mess}^5}{M_p^4\pi}F,
\end{equation}

\noindent Note that instead of a $m_{mess}^3/M_p^2$ suppression, we now have $m_{mess}^5/M_p^4$. This gives a viable scenario since the messenger mass now needed is $m_{mess}\sim 4000$ TeV, for a coupling constant of order one.

\subsection{Spin-1/2 Dark Matter}

We now consider a massive DM of spin-1/2 (let us call it $\psi$) that decays into SM particles. Without a specific model, we assume $\psi$ decays dominantly via two-body decays and focus our attention to this phenomenon. This means $\psi$ must decay into one SM fermion $f$ and one SM gauge boson $G$. Since the DM candidate must be neutral, the posibilities for a two body decay of spin 1/2 DM into SM particles are $(f,G)=(\nu,Z), (l^{\pm},W^{\mp})$. For the case that $\psi$ is a Dirac fermion, the two-body decays are mediated by the effective operator $g_DG_\mu\bar{f}\gamma^\mu(1+r\gamma_5)\psi + g_D^*G^*_\mu\bar{\psi}\gamma^\mu(1+r\gamma_5)f$. The first term gives rise to $\psi$ decay while the second one is responsible for the decay of $\bar{\psi}$. Again summing over the final state spins and averaging over the spin of the decaying $\psi$, we find the decay rate to be
\begin{equation}
\label{fG}
\Gamma(\psi\to fG) = \Gamma(\bar{\psi}\to\bar{f}\bar{G}) = \frac{|g_D|^2m_\psi^3}{16\pi m_G^2}\sqrt{\lambda\left (\frac{m_G}{m_\psi},\frac{m_f}{m_\psi}\right )}\left [ \omega\left (\frac{m_G}{m_\psi},\frac{m_f}{m_\psi}\right )+r^2\omega\left (\frac{m_G}{m_\psi},-\frac{m_f}{m_\psi}\right )\right ],
\end{equation}
where $\lambda(a,b) = (1+a-b)(1-a-b)(1-a+b)(1+a+b)$ and $\omega(a,b) = (1+a-b)(1-a-b)[2a^2+(1+b)^2]$,
and $m_A$ denotes respectively the mass of particle $A$.

Now consider the case where the decay of the fermionic DM candidate comes from an operator $g_D\bar{\psi}H(1+ir_p\gamma_5)f+g_D^*\bar{f}H(1+ir_p\gamma_5)\psi$, where $H$ is the Higgs boson\footnote{We consider the Higgs in this particular case because this interaction would arise at the same or lower order than the other one we considered. With decaying DM of other spins, there would be an additional suppression from the ratio of electroweak scale over the cut-off scale.} of the SM. In this case the decay rate of $\psi \rightarrow H + \nu$ is given by

\begin{equation}
\label{Hnu}
\Gamma(\psi\to H\nu)=\Gamma(\bar{\psi}\to H\bar{\nu})=\frac{|g_D|^2m_\psi}{16\pi}\sqrt{\lambda\left(\frac{m_H}{m_{\psi}},\frac{m_f}{m_{\psi}}\right)}\left[z\left(\frac{m_f}{m_{\psi}},\frac{m_H}{m_{\psi}}\right)+r_p^2z\left(-\frac{m_f}{m_{\psi}},\frac{m_H}{m_{\psi}}\right)\right],
\end{equation}

\noindent where $z(a,b)=1+a^2-b^2+2a$.

We can now consider a simple scenario in which $r = r_p = 0$ for all the decay channels and they all have the same coupling constant $g_D$. Then the total decay rate of $\psi$ is given by summing over all the possible channels:
\begin{eqnarray}
\label{twobody}
\Gamma(\psi\to{\rm 2 \ body}) & = & 3\ \Gamma(\psi\to Z\nu) + \Gamma(\psi\to W^+e^-) + \Gamma(\psi\to W^+\mu^-) \nonumber\\
& & + \Gamma(\psi\to W^+\tau^-) + 3\ \Gamma(\psi\to H\nu) \\
& = & 126.5\ |g_D|^2 \nonumber
\end{eqnarray}
where we have picked $m_\psi\sim 200 \mathrm{GeV}$, $H\sim 100 \mathrm{GeV}$ and assumed all three generations of neutrinos have masses $\sim 1 \mathrm{eV}$. The factor 3 in Equation (\ref{twobody}) is for three generations of neutrinos. Our cosmological bound then gives us the constraint
\begin{equation}
\hbox{Dirac fermion}:\quad |g_D|\lesssim 4.0\times 10^{-23} \ (95.4 \% \textrm{ confidence level}).
\end{equation}

For the case that $\psi$ is a Majorana fermion, the above analysis follows through. Given the same interaction terms as those shown above, the partial decay rates of a Majorana $\psi$ are exactly the same as Equations (\ref{fG}) and (\ref{Hnu}). There are, however, no distinction between $\psi$ and $\bar{\psi}$ in this case any more. In a four-component spinor notation, $\psi$ and $\bar{\psi}$ relate to each other via the charge conjugation matrix. This means the total decay rate of a Majorana $\psi$ has contributions from decays into `particles' and decays into `anti-particles'.

In the same simple scenario we considered above, the total decay rate of a Majorana $\psi$ will be increased by a factor of 2 compared to Equation (\ref{twobody}). The constraint on the coupling constant will correspondingly be tightened by a factor of $\sqrt{2}$:
\begin{equation}
\hbox{Majorana fermion}:\quad |g_M|\lesssim 2.8\times 10^{-23} \ (95.4 \% \textrm{ confidence level}).
\end{equation}

\subsubsection{R-parity violation in minimal supergravity models}

Undoubtedly, the most thoroughly investigated models in the supersymmetric menagerie are the minimal supergravity (mSUGRA) models \cite{Chamseddine:1982jx, Barbieri:1982eh, Hall:1983iz, Cremmer:1982vy, Ohta:1982wn}. While the theoretical motivation for universality of scalar masses, gaugino masses and trilinear terms is questionable (since these values depend on the mechanism by which supersymmetry breaking is transmitted to our sector), it has nevertheless remained a useful benchmark. For our purposes, it is sufficient for us to use the fact that in a variety of these mSUGRA models, the lightest supersymmetric particle (LSP) is a neutral particle that is overwhelmingly composed of the spin-1/2 supersymmetric partner of the B-gauge boson called the bino, $\tilde{B}$. There are of course technically natural classes of models \cite{Goh:2003yr, Goh:2005be, Ng:2007sm} very similar to mSUGRA theories that will give bino as the LSP, and the analysis below would similarly apply to them.


In the presence of R-parity violation, the bino LSP would of course decay. Traditionally, theories with R-parity violation were often assumed to be unable to provide a dark matter candidate. Here, we can turn this around and ask what the couplings of the theory have to be so that the theory can still furnish  us with viable dark matter candidate. To do that, we need to first explore the possible decays.

While the two-body decay might seem to have a more favorable phase space, these decays however would arise from Feynman diagrams \cite{Dawson:1985vr} only if we have the R-parity violating terms together with the introduction of an additional loop and further suppression by dimensionless ratios of electroweak scale over the cut-off scale. We will therefore assume that the bino will dominantly decay into three SM particles via the trilinear R-parity violating terms. If for some particular models, one needs to add in some of the two-body decay terms, one can look up the Appendix or the previous subsection for the relevant cross-sections and include them in the overall analysis.

 Neglecting all final state masses, the decay rate for a three-body bino decay is given by
\begin{equation}
\Gamma=\frac{1}{64\pi^3m_{\tilde{B}}}\int^{\frac{1}{2}m_{\tilde{B}}}_0dE_1\int^{\frac{1}{2}m_{\tilde{B}}}_{\frac{1}{2}m_{\tilde{B}}-E_1}dE_2 \ {\sum_{\rm spins}}|{\cal M}|^2
\end{equation}
where $E_i$ is the energy of the final particle $i$\footnote{Of course the identification of a final particle as particle $i$ is arbitrary. This arbitrariness does not change the final expression for $\Gamma$.}, and the summation symbol means averaging over initial spins and summing over final spins. The amplitudes squared for three-body decays of neutralino due to trilinear R-parity violating terms have been evaluated and shown in \cite{Dreiner:1994tj, Baltz:1997gd, Barbier:2004ez}, with the appropriate spin summing/averaging. Strictly speaking a neutralino is a superposition of the bino and three other fermionic supersymmetric particles but for our purposes, it is sufficient to consider the bino to the lightest neutralino and the LSP. The results in \cite{Baltz:1997gd} can be easily applied to LSP decay by demanding the neutralino has a 100\% bino component, i.e. by setting $N_{\chi 1} = 1$ and $N_{\chi n}=0, n=2,3,4$ in the notation of \cite{Baltz:1997gd}. For simplicity, all final state masses are neglected in the analysis below. We have also ignored the mixings and the widths of the sfermions, which mediate the decay as internal lines in the Feynman diagrams.

Given the R-parity violating superpotential term
\begin{equation}
W_{LLE} = \epsilon^{\sigma\rho}\lambda_{ijk}L_{i\sigma}L_{j\rho}E^c_k
\end{equation}
(where $i, j$ and $k$, each of which runs from 1 to 3, are generation indices, $\sigma$ and $\rho$ are $SU(2)_L$ indices, and the superscript $c$ indicates charge conjugation), the decay channel $\tilde{B} \to e^+_i\bar{\nu}_je^-_k$ is possible. Using the generic expression for amplitude squared in \cite{Baltz:1997gd} and putting in our simplifications, we get the  decay rate
\begin{eqnarray}
\Gamma(\tilde{B} \to e^+_i\bar{\nu}_je^-_k) & = & \frac{8}{128\pi^3}|\lambda_{ijk}|^2g'^2m_{\tilde{B}}\biggl [2Y^2_LK\left(\frac{m_{\tilde{e}_i}}{m_{\tilde{B}}}\right ) + 2Y^2_LK\left (\frac{m_{\tilde{\nu}_j}}{m_{\tilde{B}}}\right ) \nonumber\\
& & +\ 2Y^2_EK\left (\frac{m_{\tilde{e}_k}}{m_{\tilde{B}}}\right ) - 2Y^2_LP\left (\frac{m_{\tilde{\nu}_j}}{m_{\tilde{B}}},\frac{m_{\tilde{e}_i}}{m_{\tilde{B}}}\right ) + 2Y_LY_EP\left (\frac{m_{\tilde{e}_k}}{m_{\tilde{B}}},\frac{m_{\tilde{e}_i}}{m_{\tilde{B}}}\right ) \nonumber \\
& &+\ 2Y_LY_EP\left (\frac{m_{\tilde{e}_k}}{m_{\tilde{B}}},\frac{m_{\tilde{\nu}_j}}{m_{\tilde{B}}}\right )\biggr ]
\end{eqnarray}
where $g'$ is the gauge coupling of $U(1)_Y$, $m_{\tilde{f}_n}$ is the mass of the scalar superpartner of particle $\tilde{f}_n$ and $Y_S$ denotes the hypercharge of a superfield $S$ (for example, $Y_E=-1$). $K(x)$ and $P(x,y)$ are functions defined in the Appendix.

Another trilinear R-parity violating superpotential term is
\begin{equation}
W_{LQD} = \epsilon^{\sigma\rho}\lambda'_{ijk}L_{i\sigma}Q_{j\rho\alpha}D^c_{k\alpha}
\end{equation}
with the $SU(3)_c$ index $\alpha$. This term gives rise to the decays $\tilde{B} \to e^+_i\bar{u}_jd_k$ and $\tilde{B} \to \bar{\nu}_i\bar{d}_jd_k$. The decay rates for these channels are similar to the one above, with the appropriate substitution of superpartner masses and prefactors:
\begin{eqnarray}
\label{decay_eud}
\Gamma(\tilde{B} \to  e^+_i\bar{u}_jd_k) & = & \frac{6}{128\pi^3}|\lambda'_{ijk}|^2g'^2m_{\tilde{B}}\biggl [2Y^2_LK\left(\frac{m_{\tilde{e}_i}}{m_{\tilde{B}}}\right ) + 2Y^2_QK\left (\frac{m_{\tilde{u}_j}}{m_{\tilde{B}}}\right ) \nonumber\\
& & +\ 2Y^2_DK\left (\frac{m_{\tilde{d}_k}}{m_{\tilde{B}}}\right ) - 2Y_LY_QP\left (\frac{m_{\tilde{u}_j}}{m_{\tilde{B}}},\frac{m_{\tilde{e}_i}}{m_{\tilde{B}}}\right ) + 2Y_LY_DP\left (\frac{m_{\tilde{d}_k}}{m_{\tilde{B}}},\frac{m_{\tilde{e}_i}}{m_{\tilde{B}}}\right ) \nonumber \\
& &+\ 2Y_QY_DP\left (\frac{m_{\tilde{d}_k}}{m_{\tilde{B}}},\frac{m_{\tilde{u}_j}}{m_{\tilde{B}}}\right )\biggr ]
\end{eqnarray}
and
\begin{eqnarray}
\label{decay_nudd}
\Gamma(\tilde{B} \to  \bar{\nu}_i\bar{d}_jd_k) & = & \frac{6}{128\pi^3}|\lambda'_{ijk}|^2g'^2m_{\tilde{B}}\biggl [2Y^2_LK\left(\frac{m_{\tilde{\nu}_i}}{m_{\tilde{B}}}\right ) + 2Y^2_QK\left (\frac{m_{\tilde{d}_j}}{m_{\tilde{B}}}\right ) \nonumber\\
& & +\ 2Y^2_DK\left (\frac{m_{\tilde{d}_k}}{m_{\tilde{B}}}\right ) - 2Y_LY_QP\left (\frac{m_{\tilde{d}_j}}{m_{\tilde{B}}},\frac{m_{\tilde{\nu}_i}}{m_{\tilde{B}}}\right ) + 2Y_LY_DP\left (\frac{m_{\tilde{d}_k}}{m_{\tilde{B}}},\frac{m_{\tilde{\nu}_i}}{m_{\tilde{B}}}\right ) \nonumber \\
& &+\ 2Y_QY_DP\left (\frac{m_{\tilde{d}_k}}{m_{\tilde{B}}},\frac{m_{\tilde{d}_j}}{m_{\tilde{B}}}\right )\biggr ].
\end{eqnarray}
Note that the numerical value of an $SU(3)_c$ colour factor has been included in the prefactors of Equations (\ref{decay_eud}) and (\ref{decay_nudd}).

In a similar way, the decay channel $\tilde{B} \to  \bar{u}_i\bar{d}_j\bar{d}_k$ is allowed by the superpotential term
\begin{equation}
W_{UDD} = \epsilon^{\alpha\beta\gamma}\lambda''_{ijk}U^c_{i\alpha}D^c_{j\beta}D^c_{k\gamma},
\end{equation}
where $\alpha$, $\beta$ and $\gamma$ are all $SU(3)_c$ indices. The corresponding decay rate is
\begin{eqnarray}
\Gamma(\tilde{B} \to  \bar{u}_i\bar{d}_j\bar{d}_k) & = & \frac{48}{128\pi^3}|\lambda''_{ijk}|^2g'^2m_{\tilde{B}}\biggl [2Y^2_UK\left(\frac{m_{\tilde{u}_i}}{m_{\tilde{B}}}\right ) + 2Y^2_DK\left (\frac{m_{\tilde{d}_j}}{m_{\tilde{B}}}\right ) \nonumber\\
& & +\ 2Y^2_DK\left (\frac{m_{\tilde{d}_k}}{m_{\tilde{B}}}\right ) - 2Y_UY_DP\left (\frac{m_{\tilde{d}_j}}{m_{\tilde{B}}},\frac{m_{\tilde{u}_i}}{m_{\tilde{B}}}\right ) - 2Y_UY_DP\left (\frac{m_{\tilde{d}_k}}{m_{\tilde{B}}},\frac{m_{\tilde{u}_i}}{m_{\tilde{B}}}\right ) \nonumber \\
& &-\ 2Y_D^2P\left (\frac{m_{\tilde{d}_k}}{m_{\tilde{B}}},\frac{m_{\tilde{d}_j}}{m_{\tilde{B}}}\right )\biggr ].
\end{eqnarray}
Here a different $SU(3)_c$ colour factor has been included in the prefactor.

It should be pointed out that the bino is a Majorana fermion. This means what we shown above is only half of its possible decay channels: the other decay channels are obtained by applying charge conjugation to all the final particles in any of the above channels. The decay rates, however, are invariant under charge conjugation.

Because the $LLE$ term contains two copies of $L$'s and they contract with the Levi-Civita tensor, $\lambda_{ijk}$ is anti-symmetric in $i$ and $j$. Thus it only represents nine couplings. Similarly, $\lambda''_{ijk}$ is anti-symmetric in $j$ and $k$. This argument is not applicable to $\lambda'_{ijk}$, so it does indeed contain 27 couplings (see, for example, \cite{Dreiner:1994tj, Baltz:1997gd, Barbier:2004ez}).

As a simple application of our cosmological constraint on the DM decay rate, we assume $m_{\tilde{B}}\sim 100 \mathrm{GeV}$ and all the sfermions have masses $\sim 300 \mathrm{GeV}$. We also assume all the non-zero R-parity violating couplings attain the same value $\lambda$, i.e.
\begin{equation}
\lambda_{i_1j_1k_1} = \lambda'_{i_2j_2k_2} = \lambda''_{i_3j_3k_3} = \lambda, \quad i_1\ne j_1, j_3\ne k_3.
\end{equation}
Summing over all the possible 3-body decay channels of bino, the total decay rate is given by
\begin{eqnarray}
\Gamma(\tilde{B}\to{\rm 3 \ body}) & = & 2 \ [  \ 9\ \Gamma(\tilde{B} \to e^+_i\bar{\nu}_je^-_k)+27\ \Gamma(\tilde{B} \to  e^+_i\bar{u}_jd_k) \nonumber \\
& &+27\ \Gamma(\tilde{B} \to  \bar{\nu}_i\bar{d}_jd_k)+9\ \Gamma(\tilde{B} \to  \bar{u}_i\bar{d}_j\bar{d}_k)\ ] \\
& = & 0.00144|\lambda|^2 \nonumber
\end{eqnarray}
where we have used $g'=0.36$. Our cosmological bound then constrain the coupling constant to be
\begin{equation}
|\lambda| \lesssim 1.2\times 10^{-20} \ (95.4 \% \textrm{ confidence level}).
\end{equation}

In comparison, one of the strongest constraint on R-parity violation comes from the consequent baryon number violation that arises due to the former. Ref.\cite{Chemtob:2004xr} gives a value of $\lambda'' \lesssim 10^{-9}$ for the most constrained of all the $\lambda$'s. So if indeed the assumption that we have bino-like DM holds true, then the most stringent limits on R-parity violation would come from our analysis.

\subsection{Spin-1 Dark Matter}
We now consider a massive DM of spin-1 (let us call it $\chi$) that decays into SM particles. Since the $\chi^\mu$ field carries one Lorentz index, it contracts with other SM fields differently from the spin-0 DM, thus giving rise to different interaction terms and decay rates.

In contrast to the decay of spin-0 DM, helicity suppression is not observed in the decay of $\chi\to\nu\bar{\nu}$. $\chi^\mu$ can directly coupled to the neutrino current $\bar{\nu}_L\gamma_\mu\nu_L$, without any insertion of $\partial^\mu$. On the other hand, every spin-1 particle has to obey the Landau-Yang theorem \cite{Landau:1948, Yang:1950rg} which states that because of rotational invariance, it cannot decay into two massless spin-1 particles. Hence, the decays $\chi\to\gamma\gamma$ and $\chi\to gg$ are not allowed. The possible partial decay widths (with  summing over final state spins and averaging over the initial state spin) for a spin-1 DM are rather numerous and not that illuminating to list them all here. So we have relegated them to Table \ref{table:spin1} in the Appendix. In the case where the DM is indeed an additional U(1) gauge field that is massive, the possibility of kinetic mixing with the photon \cite{Holdom:1985ag} must be considered. Such a term could be radiatively generated via exchange of a field that is charged under both U(1)'s. Following Refs.\cite{Babu:1996vt, Babu:1997st}, we can manipulate the Lagrangian into a form where the mixing manifests itself in the coefficients of the following terms, $g_1\chi_\mu\bar{f}\gamma^\mu(1+r\gamma_5)f$. But this is a term that has already been considered in Table \ref{table:spin1}.

To get a feel for the numbers involved, let us now consider a simple model where all the interaction terms in Table \ref{table:spin1} exist, with all the coupling constants real and of the same order of magnitude. Again, we will also assume $m_{decay \ product}/m_\chi \ll 1$. Then
\begin{equation}
\Gamma(\chi \to Z\gamma)+\Gamma(\chi \to ZZ)+\Gamma(\chi \to W^+W^-)\approx\frac{g_1^2m_\chi^3}{96\pi}\left (\frac{5}{m_Z^2} + \frac{2}{m_Z^2} + \frac{4}{m_W^2}\right ).
\end{equation}
Because $m_f/m_\chi$ is small for the value of $m_\chi$ we are considering,
\begin{eqnarray}
\sum_{f\in\ {\rm SM}}\Gamma(\chi\to f\bar f) &\approx& \frac{g_1^2m_\chi}{12\pi}(1+r^2)\sum_{f\in\ {\rm SM}}N_f\nonumber \\
&=& \frac{7g_1^2m_\chi}{4\pi}(1+r^2).
\end{eqnarray}
Similar to the case of spin-0 DM, the decay to $t\bar t$ is not included here, and the parameter $r$ is also assumed to be the same for all $f$. Note also that we have different $m_\chi$ dependence for the decays into fermion-antifermion and massive gauge bosons, unlike in the spin-0 case.

Our result then gives an upper bound on the sum of all the decay rates into SM particles. For illustration, we consider $m_\chi\sim 100 \mathrm{GeV}$ and $r=0$. Our bound on $\Gamma$ can then be translated into a constraint on the coupling constant:
\begin{equation}
g_1\lesssim 5.8\times 10^{-23} \ (95.4 \% \textrm{ confidence level}),
\end{equation}
where we have substituted $m_W = 80\mathrm{GeV}$ and $m_Z = 91 \mathrm{GeV}$.

\subsubsection{T-Parity violation in little Higgs models}
Little Higgs models with T-parity violation is another possible scenario in which the dark matter candidate decays. Analogous to R-parity in SUSY models, all non-SM particles in Little Higgs model are assigned to be T-odd, while all SM ones T-even. The T-parity then requires all coupling terms to have an even number of non-SM fields. This forbids the contribution of the non-SM particles to the oblique electroweak parameters, and consequently, the symmetry breaking scale $f$ can be lowered to about 1TeV \cite{Cheng:2004yc}. The Lightest T-odd Particle (LTOP), moreover, is stable and has often been nominated as a dark matter candidate.

However, Ref.\cite{Hill:2007zv} has pointed out that anomalies in general give rise to a Wess-Zumino-Witten (WZW) term, which breaks the T-parity (Refs.\cite{Csaki:2008se, Krohn:2008ye} have constructed Little Higgs models free of the usual WZW term). This means the LTOP is not exactly stable. Indeed, phenomenological consequences of the WZW term in the Littlest Higgs model have been studied in Ref.\cite{Barger:2007df, Freitas:2008mq}. In their model, the LTOP is the massive partner of photons (denoted by $A_H$) and the WZW term contains direct couplings of $A_H$ to the Standard Model $W$ bosons, $Z$ bosons and photons. Ref.\cite{Freitas:2008mq}, moreover, pointed out that such couplings can generate two-body decay of $A_H$ to SM fermions, $A_H\to f\bar{f}$, via triangular loop diagrams.

In an attempt to be as model-independent as possible, we parameterize the couplings of $A_H$ to the SM gauge bosons as
\begin{equation}
\label{LTOP_int}
L\supset -\frac{g'}{f^2}\epsilon_{\mu\nu\rho\sigma}A_H^\mu[N_Zm_Z^2Z^\nu \partial^\rho Z^\sigma
+  N_Wm_W^2(W^{+\nu}\partial^\rho W^{-\sigma} + W^{-\nu}\partial^\rho W^{+\sigma})+ N_{AZ}m_Z^2Z^\nu F^{\rho\sigma}],
\end{equation}
\noindent where $f$ is the symmetry breaking scale, $g'$ the $U(1)$ gauge coupling, $N_Z$, $N_W$ and $N_{AZ}$ are numbers whose values depend on the exact realization and the UV completion.

Generically, the mass of $A_H$ is proportional to $f$.  If we take $f$ to be the natural symmetry breaking scale (i.e. $\sim 1TeV$) in Little Higgs models, then $m_{A_H}\gtrsim 2m_Z$. As an example, in \cite{Barger:2007df, Freitas:2008mq}, we have
\begin{equation}
m_{A_H} = \frac{g'f}{\sqrt{5}}\left [1-\frac{5v^2}{8f^2}+{\cal O} \left (\frac{v^4}{f^4}\right )\right ],
\end{equation}
\noindent where $v=246 \mathrm{GeV}$ is the Higgs vev. The condition for $m_{A_H}\gtrsim 2m_Z$ is satisfied when $g'\sim 0.36$ and $f\gtrsim 1165 \mathrm{GeV}$.

In the case of $m_{A_H}\gtrsim 2m_Z$, the decay channels of $A_H\to ZZ$ and $A_H\to W^+W^-$ are kinematically allowed, and, for simplicity, we assume these processes (together with $A_H\to Z\gamma$) to be the dominant ones. With the interaction terms in equation (\ref{LTOP_int}), the decay rates for these channels at the lowest order are given by
\begin{eqnarray}
\Gamma(A_H \to ZZ) &=& \frac{g'^2N_Z^2m_{A_H}^3m_Z^2}{96\pi f^4}\left ( 1-4\frac{m_Z^2}{m_{A_H}^2}\right )^\frac{5}{2}, \\
\Gamma(A_H \to W^+W^-) &=& \frac{g'^2N_W^2m_{A_H}^3m_W^2}{48\pi f^4}\left ( 1 - \frac{m_W^2}{m_{A_H}^2}\right )^\frac{5}{2}, \\
\Gamma(A_H \to Z\gamma) & = & \frac{g'^2N_{AZ}^2m_{A_H}^3m_Z^2}{24\pi f^4}\left ( 1+\frac{m_Z^2}{m_{A_H}^2}\right )\left ( 1-\frac{m_Z^2}{m_{A_H}^2}\right )^3.
\end{eqnarray}
\noindent The sum of these decay rates is then constrained by our bound on the dark matter lifetime. For the Littlest Higgs model, $N_{AZ} = 0$ and $N_W=N_Z$. The sum of the above decay rates is then reduced to
\begin{equation}
\Gamma = \frac{g'^5N_Z^2m_Z^2}{160\sqrt{5}\pi f},
\end{equation}
\noindent where we have used the approximations $m_W\simeq m_Z$, $m_{A_H} \simeq g'f/\sqrt{5}$ and have neglected all the mass ratios. Our bound on $\Gamma$ then gives us the constraint
\begin{equation}
\frac{N_Z^2}{f} < 4.7\times 10^{-42} \mathrm{GeV} \,(95.4 \% \textrm{ confidence level}),
\end{equation}
\noindent which of course is not reasonable as we typically expect $N_Z \sim 1$ and $f \sim 1$ TeV. But it vividly illustrates the utility of our approach when it comes to ruling out particle physics models that claim to have dark matter candidates.


\subsection{General Dimensional Considerations}

We can draw some generalizations from the above cases if we do a simple dimensional analysis.  The coupling of a spin-0, spin-1/2 or spin-1 dark matter candidate $S$ to an operator $O$ can be parameterized, with suppression of  indices and $\mathcal{O}(1)$ factors, as
\begin{equation}
L\supset g\frac{S}{\Lambda^{n-4}}O,
\end{equation}
where $g$ is a dimensionless coupling constant, $\Lambda$ is the scale where unknown new physics is integrated out to give the operator $O$, and $n$ is the sum of the dimensions of $S$ and $O$. The decay rate for such dark matter candidate is given in general by
\begin{equation}
\Gamma = \frac{g^2}{\Lambda^{2n-8}}m_S^{2n-7} F_k,
\end{equation}
where $F_k$ is a function that contains the kinematics of the decay, assumed to be of order one for simplicity.
For $n=5$, $m_S=100 \mathrm{GeV}$ and $g\sim{\cal O}(1)$, the cutoff scale should be of the order of $\Lambda \gtrsim  10^{24} \mathrm{GeV}$, suggesting that we must go to operators of higher dimensions and thus more $\Lambda$ suppression. For $n=6$, $m_S=100 \mathrm{GeV}$ and $g\sim{\cal O}(1)$, the cutoff scale can be as low as $\Lambda \sim  10^{13} \mathrm{GeV}$. On the other hand, for $n=5$, if the cutoff is taken at the Planck scale ($\Lambda\sim
10^{19} \mathrm{GeV}$) and we keep the same value of $m_S$, the coupling constant can only be as large as $g \sim 10^{-5}$. Finally to recover the cases discussed above for spin-0, spin-1/2 and spin-1 particles, we can take $n=4$ and $m_S \sim 100$ GeV to give us a coupling constant as large as $g \sim 10^{-22}$.

A few words should be reiterated about the smallness of the coupling constant. We had taken an extremely conservative value for our cutoffs, usually $\sim 10$ TeV. In an effective theory with a low cutoff arising from a high scale fundamental theory, say at Planck scale, there will be a multitude of effective operators containing mass insertions (leading to small dimensionless ratios such as $m/ M_P$) or loops (giving factors of $1/16 \pi^2$), making these tiny coupling constants natural. The small dimensionless ratios could arise from, say, the decay being mediated by some massive field much like what we have in proton decay via exchange of heavy X bosons in the context of Grand Unified Theories. The onus is then on the model builders to refine their models in a technically natural way to satisfy the constraints we have derived above without having to compromise other phenomenological constraints on their models.

\section{Conclusions and Outlook}

We have performed a full cosmological analysis using the available datasets from cosmic microwave background, Type Ia supernova, Lyman-$\alpha$ forest, galaxy clustering and weak lensing observations to determine the extent by which we can constrain decaying dark matter models which are very typical in most extensions of the Standard Model of particle physics.

In the scenario where there is negligible reionization of the baryonic gas by the decaying dark matter, we have found that the late-time Integrated Sachs-Wolfe  effect gives the strongest constraint. The lifetime of a decaying dark matter has the bound $\Gamma^{-1}\gtrsim 100$Gyr (at 95.4\% confidence level). Because of cosmic variance, the results are not likely to improve significantly with the WMAP-9yr data.

When there is significant reionization of the baryonic gas due to the decaying dark matter, the bounds become more restrictive as the CMB polarization is well measured. In this scenario, the lifetime of a decaying dark matter is $(f \, \Gamma) ^{-1}\gtrsim 5.3 \times 10^8$ Gyr (at 95.4\% confidence level) where $f$ is a phenomenological factor related to the degree of reionization. With even more CMB polarization data, one could conceivably distinguish the reionization due to decaying dark matter from reionization due to star formation, thereby giving us even better bounds on the lifetime of the dark matter. We expect that the the 21cm cosmological observation in the future would give us even greater precision as it is expected to probe the reionization history at redshifts $6<z<30$.

Having obtained the cosmological constraints, we turned our attentions to the particle physics aspects of it. For completeness and motivated by the utility of such an exercise, we systematically tabulated the decay cross-sections for a spin-0, spin-1/2 and spin-1 dark matter candidate into the Standard Model degrees of freedom. This enabled us to simply sum up all the relevant contributions for a particular model of particle physics and arrive at the functional form of the lifetime of the decaying dark matter. We repeated this process for a variety of representative models from the following classes of theories: generic supersysmmetric scenario, gauge-mediated supersymmetry breaking models and the little Higgs theories. Imposing the limits from our cosmological analysis, we find that generically for most models we have looked at, the dimensionless coupling for a decaying dark matter to Standard Model fields should be smaller than $10^{-22}$.

This restriction can be slightly relaxed if the dark matter decays solely into light particles via helicity suppressed interaction terms, in which case, the small mass of the decay products suppresses the decay rate. If, for instance, the dark matter decays purely via helicity suppressed terms into $\nu\bar{\nu}$ with Dirac mass of $\sim 2 \textrm{eV}$, then the dimensionless coupling can be as large as $10^{-11}$. In addition to constraining the coupling, one can assume it to be of ${\cal O}(1)$ and estimate the scale of new physics which suppresses the decay rate. In all cases, either the coupling attains a small value or the new physics come from a huge scale, both of which would need interesting and exotic physics to realize if indeed the dark matter does decay via dimension-4 or dimension-5 terms. In the case of exclusive helicity suppressed decays, moreover, one has to explain why other interaction terms are absent in the model. A more promising avenue, which we briefly mentioned in the previous section, is to look at models where the dark matter decays via dimension-6 operators. The Large Hadron Collider might provide us with the identity for dark matter in the very near future, but on the basis of our analysis, there will still be much to understand about physics of the dark matter sector and how it interacts with the Standard Model.

In a future paper, we hope to address some of the astrophysical issues of decaying dark matter. The recent spate of results from astrophysical experiments \cite{Strong:2005zx,Aguilar:2007yf,Bernabei:2008yi,Adriani:2008zr} has given us much to ponder. The immediate goal would of course be to combine all the astrophysical datasets with the cosmological ones that we have considered in the present paper and arrive at a set of characteristics that a phenomenologically viable decaying dark matter must possess. However, we expect considerable tension between the two classes of constraints. The astrophysical ones require that the decays to be significant enough to account for the as-yet-unexplained phenomena, while the cosmological ones need decays to be small enough because of the late time ISW effect and the CMB polarization observations. To reconcile and resolve these two seemingly conflicting classes of observations could be the defining challenge of dark matter physics in the next decade.

\begin{acknowledgments}
The work of  S.D.A., W.M.Y.C., Z.H. and S.P.N. were supported by the Natural Sciences and Engineering Research Council of Canada. S.D.A. acknowledges the support of the Direcci\'on General de Relaciones Internacionales de la Secretar\' ia de Educaci\' on P\'ublica (DGRI-SEP) of Mexico while S.P.N. was also partially supported by the Ontario Premier's Research Excellence Award. The authors would like to acknowledge Andrew Blechman, Dick Bond, Patrick Fox, Hock-Seng Goh, Bob Holdom,  Kiyotomo Ichiki, Lev Kofman, Axel Krause, Massimiliano Lattanzi, Michael Luke, Pat McDonald, Christoph Pfrommer, Dmitri Pogosyan, Erich Poppitz, Dominik Schleicher, Pascal Vaudrevange and Scott Watson for useful discussions. The authors would also like to record their thanks to Erich Poppitz and Scott Watson for careful reading of the manuscript.
\end{acknowledgments}

\newpage

\appendix*
\section{Compendium of Decay Rates}

This appendix summarizes the lowest order decay rates due to various generic interaction terms in the Lagrangian, averaging over the spin of the DM and summing over the spins of the decay products. Tables \ref{table:spin0}, \ref{table:spin1/2} and \ref{table:spin1} respectively tabulate the decay of a spin-0 DM particle $S$, a spin-1/2 DM particle $\psi$ and a spin-1 DM particle $\chi$, into SM particles. In a model-independent way, we write down generic Lagrangian terms which describe possible decay channels of DM particles to SM particles. The exact mechanisms which mediate these decays are captured by the dimensionless coupling constants, $g_0$, $g_D$ and $g_1$. The reality of the interaction terms requires the coupling constants to be real, except in the case of $\chi\to W^+W^-$ and the decay of $\psi$, in which a complex coupling constant is possible.

We follow standard conventions in denoting our fields.  $Y^\mu$ represents a gauge field while  $Y^{\mu\nu}$ is the corresponding field strength tensor. For various decay channels, $N_f$ and $N_g$ respectively denote the number of colours of a fermion species $f$ and a gluon $g$. $r$, $r_p$ and $r_a$ are parameters that describe the relative size of two interaction terms.

\begin{table}[h]
 \centering
\caption{Decay Rate of Spin-0 DM via Different Interaction Terms}\label{spin0decay}
 \label{table:spin0}
 \begin{tabular}{|c|ccl|}
 \hline
 Interaction Term &  & & Decay Rate \\
 \hline
 $g_0S\bar{f}(1+ir_p\gamma_5)f$ & $\Gamma(S \to f \bar{f})$ & = &$\frac{g_0^2m_S N_f}{4\pi}\sqrt{1-4\frac{m_f^2}{m_S^2}}\times \left (\frac{1+r_p^2}{2}-2\frac{m_f^2}{m_S^2}\right )$ \\
 \hline
 $\frac{g_0}{\Lambda}\bar{f}\gamma^\mu(1+r_a\gamma_5)f\partial_\mu S$ & $\Gamma(S \to f \bar{f})$ & = &$\frac{g_0^2r_a^2 N_f}{2\pi}m_f\frac{m_fm_S}{\Lambda^2}\sqrt{1-4\frac{m_f^2}{m_S^2}}$ \\
 \hline
 $\frac{g_{0s}}{\Lambda}SF_{\mu\nu}F^{\mu\nu}$ & $\Gamma(S \to \gamma\gamma)$ & = & $\frac{g_{0s}^2m_S^3}{4\pi\Lambda^2}$ \\
 \hline
 $\frac{g_{0p}}{\Lambda}S\epsilon_{\mu\nu\sigma\lambda}F^{\mu\nu}F^{\sigma\lambda}$ & $\Gamma(S \to \gamma\gamma)$ & = & $\frac{g_{0p}^2}{\pi\Lambda^2}m_S^3$ \\
 \hline
 $\frac{g_{0s}}{\Lambda}SG^a_{\mu\nu}G^{a,\mu\nu}$ & $\Gamma(S \to gg)$ & = & $\frac{g_{0s}^2m_S^3 N_g}{4\pi\Lambda^2}$ \\
 \hline
 $\frac{g_{0p}}{\Lambda}S\epsilon_{\mu\nu\sigma\lambda}G^{a,\mu\nu}G^{a,\sigma\lambda}$ & $\Gamma(S \to gg)$ & = & $\frac{g_{0p}^2}{\pi\Lambda^2}N_g m_S^3$ \\
 \hline
  $\frac{g_0m_Z^2}{\Lambda}SZ_\mu Z^\mu$ & $\Gamma(S \to ZZ)$ & = & $\frac{g_{0}^2m_S^3}{32\pi\Lambda^2}\sqrt{1-4\frac{m_Z^2}{m_S^2}}\times \left (1-4\frac{m_Z^2}{m_S^2}+12\frac{m_Z^4}{m_S^4}\right )$ \\
 \hline
 $\frac{g_{0s}}{\Lambda}SZ_{\mu\nu}Z^{\mu\nu}$ & $\Gamma(S \to ZZ)$ & = & $\frac{g_{0s}^2m_S^3}{4\pi\Lambda^2}\sqrt{1-4\frac{m_Z^2}{m_S^2}}\times \left (1-4\frac{m_Z^2}{m_S^2}+6\frac{m_Z^4}{m_S^4}\right )$ \\
 \hline
 $\frac{g_{0p}}{\Lambda}S\epsilon_{\mu\nu\sigma\lambda}Z^{\mu\nu}Z^{\sigma\lambda}$ & $\Gamma(S \to ZZ)$ & = & $\frac{g_{0p}^2m_S^3}{\pi\Lambda^2}\left (1-4\frac{m_Z^2}{m_S^2}\right )^\frac{3}{2}$ \\
 \hline
 $\frac{g_0m_W^2}{\Lambda}SW_\mu^+W^{-\mu}$ & $\Gamma(S \to W^+W^-)$ & = & $\frac{g_{0}^2m_S}{64\pi\Lambda^2}\sqrt{1-4\frac{m_W^2}{m_S^2}}\times \left (1-4\frac{m_W^2}{m_S^2}+12\frac{m_W^4}{m_S^4}\right )$ \\
 \hline
 $\frac{g_{0s}}{\Lambda}SW^+_{\mu\nu}W^{-\mu\nu}$ & $\Gamma(S \to W^+W^-)$ & = & $ \frac{g_{0s}^2m_S^3}{4\pi\Lambda^2}\sqrt{1-4\frac{m_W^2}{m_S^2}}\times \left (1-4\frac{m_W^2}{m_S^2}+6\frac{m_W^4}{m_S^4}\right )$ \\
 \hline
 $\frac{g_{0p}}{\Lambda}S\epsilon_{\mu\nu\sigma\lambda}W^{+\mu\nu}W^{-\sigma\lambda}$ & $\Gamma(S \to W^+W^-)$ & = & $\frac{g_{0p}^2m_S^3}{\pi\Lambda^2}\left (1-4\frac{m_W^2}{m_S^2}\right )^\frac{3}{2}$ \\
 \hline
 $\frac{g_{0s}}{\Lambda}F^{\mu\nu}Z_\mu\partial_\nu S$ & $\Gamma(S \to Z\gamma)$ & = & $\frac{g_{0s}^2m_S^3}{32\pi \Lambda^2}\left (1-\frac{m_Z^2}{m_S^2}\right )^3$ \\
 \hline
 $\frac{g_{0p}}{\Lambda}\epsilon_{\mu\nu\sigma\lambda}F^{\mu\nu}Z^\sigma\partial^\lambda S$ & $\Gamma(S \to Z\gamma)$ & = & $\frac{g_{0p}^2m_S^3}{8\pi\Lambda^2}\left (1-\frac{m_Z^2}{m_S^2}\right )^3$ \\
 \hline
 \end{tabular}
\end{table}

\begin{table}[hp]
 \centering
\caption{Two-Body Decay Rate of Spin-1/2 DM via Generic Interaction Terms}\label{spin1/2decay}
 \label{table:spin1/2}
 \begin{tabular}{|c|ccl|}
 \hline
 Interaction Term &  & & Decay Rate \\
 \hline
$g_DG_\mu\bar{f}\gamma^\mu(1+r\gamma_5)\psi$ & $\Gamma(\psi\to fG)$ & = & $\Gamma(\bar{\psi}\to\bar{f}\bar{G})$ \\
$+ g_D^*G^*_\mu\bar{\psi}\gamma^\mu(1+r\gamma_5)f$ & & = & $\frac{|g_D|^2m_\psi^3}{16\pi m_G^2}\sqrt{\lambda\left (\frac{m_G}{m_\psi},\frac{m_f}{m_\psi}\right )}\left [ \omega\left (\frac{m_G}{m_\psi},\frac{m_f}{m_\psi}\right )+r^2\omega\left (\frac{m_G}{m_\psi},-\frac{m_f}{m_\psi}\right )\right ]$ \\
\hline
$g_D\bar{\psi}H(1+ir_p\gamma_5)f$ & $\Gamma(\psi\to Hf)$ & = & $\Gamma(\bar{\psi}\to H\bar{f})$ \\
$+g_D\bar{f}H(1+ir_p\gamma_5)\psi$ & & = & $\frac{g_D^2m_\psi}{16\pi}\sqrt{\lambda\left(\frac{m_H}{m_{\psi}},\frac{m_f}{m_{\psi}}\right)}\left[z\left(\frac{m_f}{m_{\psi}},\frac{m_H}{m_{\psi}}\right)+r_p^2z\left(-\frac{m_f}{m_{\psi}},\frac{m_H}{m_{\psi}}\right)\right]$ \\
\hline
 \end{tabular}
\end{table}

\begin{table}[hp]
 \centering
\caption{Functions used for the Analysis of Bino Decay}
 \label{appendix:bino_decay}
 \begin{tabular}{|c|c|}
 \hline
 & \\
$K(x)$ & $\frac{1}{16}\left[-5+6x^2+ 2(1-4x^2+3x^4) \ {\rm ln}\left (1-\frac{1}{x^2}\right )\right ]$  \\
& \\
\hline
& \\
$P(x,y)$ & $\frac{1}{24}\left[\frac{3}{2}+\left(\frac{\pi^2y^2}{2}-6\right)x^2\right]+\frac{x^2y^2}{4} \ {\rm ln}\left(x^2+y^2-1\right)\left[-\frac{1}{2} \ {\rm ln} \left(x^2+y^2-1\right)+ \ {\rm ln} \left(x^2-1\right)\right]$ \\
& \\
& $+\frac{x^2}{4}\left(x^2-1\right) \ {\rm ln} \left(\frac{x^2}{x^2-1}\right)+\frac{x^2y^2}{4} \ {\rm ln} \left(y^2\right) \ {\rm ln} \left(\frac{x}{x^2-1}\right) -\frac{x^2y^2}{4} \ {\rm Li_2} \left(\frac{x^2-1}{x^2+y^2-1}\right) + x \leftrightarrow y$ \\
& \\
\hline
\end{tabular}
\end{table}

\begin{table}[hp]
 \centering
\caption{Decay Rate of Spin-1 DM via Different Interaction Terms}\label{spin1decay}
 \label{table:spin1}
 \begin{tabular}{|c|ccl|}
 \hline
 Interaction Term &  & & Decay Rate \\
 \hline
$g_1\chi_\mu\bar{f}\gamma^\mu(1+r\gamma_5)f$ & $\Gamma(\chi \to f \bar{f})$ & = &$\frac{g_1^2 N_f}{12\pi}m_\chi\sqrt{1-4\frac{m_f^2}{m_\chi^2}}\times\left [ 1+2\frac{m_f^2}{m_\chi^2}+r^2\left ( 1-4\frac{m_f^2}{m_\chi^2}\right ) \right ]$ \\
 \hline
  $\frac{g_1}{\Lambda}\chi^{\mu} \bar{f} \partial_{\mu}f$ & $\Gamma(\chi \to f\bar{f})$ & = & $\frac{g_1^2m_{\chi}^3}{64\pi\Lambda^2}\left(1-4\frac{m_f^2}{m_{\chi}^2}\right)^{\frac{3}{2}}$ \\
  \hline
  & $\Gamma(\chi\to \gamma\gamma \ {\rm or} \ gg)$ & = & 0 \\
 & & & forbidden by the Landau-Yang theorem \\
  \hline
  $g_1Z_\mu Z^\nu\partial_\nu\chi^\mu$ & $\Gamma(\chi \to ZZ)$ & = & $\frac{g_1^2m_\chi^3}{96\pi m_Z^2}\left ( 1-4\frac{m_Z^2}{m_\chi^2}\right )^\frac{3}{2}$ \\
  \hline
  $g_1\epsilon_{\mu\nu\rho\sigma}\chi^\mu Z^\nu\partial^\sigma Z^\rho$ & $\Gamma(\chi \to ZZ)$ & = & $\frac{g_1^2m_\chi^3}{96\pi m_z^2}\left ( 1-4\frac{m_Z^2}{m_\chi^2}\right )^\frac{5}{2}$ \\
   \hline
   $g_1W^+_\mu W^{-\nu}\partial_\nu\chi^\mu$ & $\Gamma(\chi \to W^+W^-)$ & = & $\frac{m_\chi^5}{192\pi m_W^4}\left (1-4\frac{m_W^2}{m_\chi^2}\right )^\frac{3}{2}$ \\
   $+g_1^*W^-_\mu W^{+\nu}\partial_\nu\chi^\mu$ & & & $\times\left \{ 4[Re(g_1)]^2\frac{m_W^2}{m_\chi^2} + [Im(g_1)]^2\left ( 1+4\frac{m_W^2}{m_\chi^2}\right )\right \} $ \\
   \hline
   $g_1\epsilon_{\mu\nu\rho\sigma}\chi^\mu W^{+\nu}\partial^\sigma W^{-\rho}$ & $\Gamma(\chi \to W^+W^-)$ & = & $\frac{m_\chi^3}{48\pi m_W^2}\sqrt{1-4\frac{m_W^2}{m_\chi^2}}$ \\
   $+ g_1^*\epsilon_{\mu\nu\rho\sigma}\chi^\mu W^{-\nu}\partial^\sigma W^{+\rho}$ & & & $\times\left \{ [Re(g_1)]^2\left ( 1 - 4\frac{m_W^2}{m_\chi^2}\right )^2 + [Im(g_1)]^2\left ( 1+2\frac{m_W^2}{m_\chi^2}\right ) \right \}$ \\
   \hline
   $g_1\chi_\mu Z_\nu F^{\mu\nu}$ & $\Gamma(\chi \to Z\gamma)$ & = & $\frac{g_1^2m_\chi^3}{96\pi m_Z^2}\left ( 1+\frac{m_Z^2}{m_\chi^2}\right )\left ( 1-\frac{m_Z^2}{m_\chi^2}\right )^3$ \\
  \hline
  $g_1\epsilon_{\mu\nu\rho\sigma}\chi^\mu Z^\nu F^{\sigma\rho}$ & $\Gamma(\chi \to Z\gamma)$ & = & $\frac{g_1^2m_\chi^3}{24\pi m_Z^2}\left ( 1+\frac{m_Z^2}{m_\chi^2}\right )\left ( 1-\frac{m_Z^2}{m_\chi^2}\right )^3$ \\
   \hline
 \end{tabular}
\end{table}

\newpage


\bibliography{decaydm2}

\begin{thebibliography}{94}
\expandafter\ifx\csname natexlab\endcsname\relax\def\natexlab#1{#1}\fi
\expandafter\ifx\csname bibnamefont\endcsname\relax
  \def\bibnamefont#1{#1}\fi
\expandafter\ifx\csname bibfnamefont\endcsname\relax
  \def\bibfnamefont#1{#1}\fi
\expandafter\ifx\csname citenamefont\endcsname\relax
  \def\citenamefont#1{#1}\fi
\expandafter\ifx\csname url\endcsname\relax
  \def\url#1{\texttt{#1}}\fi
\expandafter\ifx\csname urlprefix\endcsname\relax\def\urlprefix{URL }\fi
\providecommand{\bibinfo}[2]{#2}
\providecommand{\eprint}[2][]{\url{#2}}

\bibitem[{\citenamefont{{Hinshaw} et~al.}(2008)\citenamefont{{Hinshaw},
  {Weiland}, {Hill}, {Odegard}, {Larson}, {Bennett}, {Dunkley}, {Gold},
  {Greason}, {Jarosik} et~al.}}]{Hinshaw2008}
\bibinfo{author}{\bibfnamefont{G.}~\bibnamefont{{Hinshaw}}},
  \bibinfo{author}{\bibfnamefont{J.~L.} \bibnamefont{{Weiland}}},
  \bibinfo{author}{\bibfnamefont{R.~S.} \bibnamefont{{Hill}}},
  \bibinfo{author}{\bibfnamefont{N.}~\bibnamefont{{Odegard}}},
  \bibinfo{author}{\bibfnamefont{D.}~\bibnamefont{{Larson}}},
  \bibinfo{author}{\bibfnamefont{C.~L.} \bibnamefont{{Bennett}}},
  \bibinfo{author}{\bibfnamefont{J.}~\bibnamefont{{Dunkley}}},
  \bibinfo{author}{\bibfnamefont{B.}~\bibnamefont{{Gold}}},
  \bibinfo{author}{\bibfnamefont{M.~R.} \bibnamefont{{Greason}}},
  \bibinfo{author}{\bibfnamefont{N.}~\bibnamefont{{Jarosik}}},
  \bibnamefont{et~al.}, \textbf{\bibinfo{volume}{803}} (\bibinfo{year}{2008}),
  \eprint{0803.0732}.

\bibitem[{\citenamefont{{Komatsu} et~al.}(2008)\citenamefont{{Komatsu},
  {Dunkley}, {Nolta}, {Bennett}, {Gold}, {Hinshaw}, {Jarosik}, {Larson},
  {Limon}, {Page} et~al.}}]{Komatsu2008}
\bibinfo{author}{\bibfnamefont{E.}~\bibnamefont{{Komatsu}}},
  \bibinfo{author}{\bibfnamefont{J.}~\bibnamefont{{Dunkley}}},
  \bibinfo{author}{\bibfnamefont{M.~R.} \bibnamefont{{Nolta}}},
  \bibinfo{author}{\bibfnamefont{C.~L.} \bibnamefont{{Bennett}}},
  \bibinfo{author}{\bibfnamefont{B.}~\bibnamefont{{Gold}}},
  \bibinfo{author}{\bibfnamefont{G.}~\bibnamefont{{Hinshaw}}},
  \bibinfo{author}{\bibfnamefont{N.}~\bibnamefont{{Jarosik}}},
  \bibinfo{author}{\bibfnamefont{D.}~\bibnamefont{{Larson}}},
  \bibinfo{author}{\bibfnamefont{M.}~\bibnamefont{{Limon}}},
  \bibinfo{author}{\bibfnamefont{L.}~\bibnamefont{{Page}}},
  \bibnamefont{et~al.}, \textbf{\bibinfo{volume}{803}} (\bibinfo{year}{2008}),
  \eprint{0803.0547}.

\bibitem[{\citenamefont{Schmaltz and Tucker-Smith}(2005)}]{Schmaltz:2005ky}
\bibinfo{author}{For a review, see \bibfnamefont{M.}~\bibnamefont{Schmaltz}} \bibnamefont{and}
  \bibinfo{author}{\bibfnamefont{D.}~\bibnamefont{Tucker-Smith}},
  \bibinfo{journal}{Ann. Rev. Nucl. Part. Sci.} \textbf{\bibinfo{volume}{55}},
  \bibinfo{pages}{229} (\bibinfo{year}{2005}), \eprint{hep-ph/0502182}.

\bibitem[{\citenamefont{Martin}(1997)}]{Martin:1997ns}
\bibinfo{author}{For a review, see \bibfnamefont{S.~P.} \bibnamefont{Martin}}
  (\bibinfo{year}{1997}), \eprint{hep-ph/9709356}.

\bibitem[{\citenamefont{Hill and Hill}(2007)}]{Hill:2007zv}
\bibinfo{author}{\bibfnamefont{C.~T.} \bibnamefont{Hill}} \bibnamefont{and}
  \bibinfo{author}{\bibfnamefont{R.~J.} \bibnamefont{Hill}},
  \bibinfo{journal}{Phys. Rev.} \textbf{\bibinfo{volume}{D76}},
  \bibinfo{pages}{115014} (\bibinfo{year}{2007}), \eprint{0705.0697}.

\bibitem[{\citenamefont{Barbier et~al.}(2005)}]{Barbier:2004ez}
\bibinfo{author}{For a review, see \bibfnamefont{R.}~\bibnamefont{Barbier}} \bibnamefont{et~al.},
  \bibinfo{journal}{Phys. Rept.} \textbf{\bibinfo{volume}{420}},
  \bibinfo{pages}{1} (\bibinfo{year}{2005}), \eprint{hep-ph/0406039}.

\bibitem[{\citenamefont{Israel}(1967)}]{b1}
\bibinfo{author}{\bibfnamefont{W.}~\bibnamefont{Israel}},
  \bibinfo{journal}{Phys. Rev.} \textbf{\bibinfo{volume}{164}},
  \bibinfo{pages}{1776} (\bibinfo{year}{1967}).

\bibitem[{\citenamefont{Israel}(1968)}]{b2}
\bibinfo{author}{\bibfnamefont{W.}~\bibnamefont{Israel}},
  \bibinfo{journal}{Commun. Math. Phys.} \textbf{\bibinfo{volume}{8}},
  \bibinfo{pages}{245} (\bibinfo{year}{1968}).

\bibitem[{\citenamefont{Carter}(1971)}]{b3}
\bibinfo{author}{\bibfnamefont{B.}~\bibnamefont{Carter}},
  \bibinfo{journal}{Phys. Rev. Lett.} \textbf{\bibinfo{volume}{26}},
  \bibinfo{pages}{331} (\bibinfo{year}{1971}).

\bibitem[{\citenamefont{Wald}(1971)}]{b4}
\bibinfo{author}{\bibfnamefont{R.~M.} \bibnamefont{Wald}},
  \bibinfo{journal}{Phys. Rev. Lett.} \textbf{\bibinfo{volume}{26}},
  \bibinfo{pages}{1653} (\bibinfo{year}{1971}).

\bibitem[{\citenamefont{Banks}(1989)}]{Banks:1989ag}
\bibinfo{author}{\bibfnamefont{T.}~\bibnamefont{Banks}},
  \bibinfo{journal}{Nucl. Phys.} \textbf{\bibinfo{volume}{B323}},
  \bibinfo{pages}{90} (\bibinfo{year}{1989}).

\bibitem[{\citenamefont{Krauss and Wilczek}(1989)}]{Krauss:1988zc}
\bibinfo{author}{\bibfnamefont{L.~M.} \bibnamefont{Krauss}} \bibnamefont{and}
  \bibinfo{author}{\bibfnamefont{F.}~\bibnamefont{Wilczek}},
  \bibinfo{journal}{Phys. Rev. Lett.} \textbf{\bibinfo{volume}{62}},
  \bibinfo{pages}{1221} (\bibinfo{year}{1989}).

\bibitem[{\citenamefont{Cen}(2000)}]{Cen:2000xv}
\bibinfo{author}{\bibfnamefont{R.}~\bibnamefont{Cen}} (\bibinfo{year}{2000}),
  \eprint{astro-ph/0005206}.

\bibitem[{\citenamefont{Borzumati et~al.}(2008)\citenamefont{Borzumati,
  Bringmann, and Ullio}}]{Borzumati:2008zz}
\bibinfo{author}{\bibfnamefont{F.}~\bibnamefont{Borzumati}},
  \bibinfo{author}{\bibfnamefont{T.}~\bibnamefont{Bringmann}},
  \bibnamefont{and} \bibinfo{author}{\bibfnamefont{P.}~\bibnamefont{Ullio}},
  \bibinfo{journal}{Phys. Rev.} \textbf{\bibinfo{volume}{D77}},
  \bibinfo{pages}{063514} (\bibinfo{year}{2008}), \eprint{hep-ph/0701007}.

\bibitem[{\citenamefont{Ichiki et~al.}(2004)\citenamefont{Ichiki, Oguri, and
  Takahashi}}]{Ichiki:2004vi}
\bibinfo{author}{\bibfnamefont{K.}~\bibnamefont{Ichiki}},
  \bibinfo{author}{\bibfnamefont{M.}~\bibnamefont{Oguri}}, \bibnamefont{and}
  \bibinfo{author}{\bibfnamefont{K.}~\bibnamefont{Takahashi}},
  \bibinfo{journal}{Phys. Rev. Lett.} \textbf{\bibinfo{volume}{93}},
  \bibinfo{pages}{071302} (\bibinfo{year}{2004}), \eprint{astro-ph/0403164}.

\bibitem[{\citenamefont{Gong and Chen}(2008)}]{Gong:2008gi}
\bibinfo{author}{\bibfnamefont{Y.}~\bibnamefont{Gong}} \bibnamefont{and}
  \bibinfo{author}{\bibfnamefont{X.}~\bibnamefont{Chen}}
  (\bibinfo{year}{2008}), \eprint{0802.2296}.

\bibitem[{\citenamefont{Lattanzi}(2008)}]{Lattanzi:2008zz}
\bibinfo{author}{\bibfnamefont{M.}~\bibnamefont{Lattanzi}},
  \bibinfo{journal}{AIP Conf. Proc.} \textbf{\bibinfo{volume}{966}},
  \bibinfo{pages}{163} (\bibinfo{year}{2008}).

\bibitem[{\citenamefont{Zhang et~al.}(2007)\citenamefont{Zhang, Chen,
  Kamionkowski, Si, and Zheng}}]{Zhang:2007zzh}
\bibinfo{author}{\bibfnamefont{L.}~\bibnamefont{Zhang}},
  \bibinfo{author}{\bibfnamefont{X.}~\bibnamefont{Chen}},
  \bibinfo{author}{\bibfnamefont{M.}~\bibnamefont{Kamionkowski}},
  \bibinfo{author}{\bibfnamefont{Z.-g.} \bibnamefont{Si}}, \bibnamefont{and}
  \bibinfo{author}{\bibfnamefont{Z.}~\bibnamefont{Zheng}},
  \bibinfo{journal}{Phys. Rev.} \textbf{\bibinfo{volume}{D76}},
  \bibinfo{pages}{061301} (\bibinfo{year}{2007}), \eprint{0704.2444}.

\bibitem[{\citenamefont{Palomares-Ruiz}(2008)}]{PalomaresRuiz:2007ry}
\bibinfo{author}{\bibfnamefont{S.}~\bibnamefont{Palomares-Ruiz}},
  \bibinfo{journal}{Phys. Lett.} \textbf{\bibinfo{volume}{B665}},
  \bibinfo{pages}{50} (\bibinfo{year}{2008}), \eprint{0712.1937}.

\bibitem[{\citenamefont{Yuksel and Kistler}(2008)}]{Yuksel:2007dr}
\bibinfo{author}{\bibfnamefont{H.}~\bibnamefont{Yuksel}} \bibnamefont{and}
  \bibinfo{author}{\bibfnamefont{M.~D.} \bibnamefont{Kistler}},
  \bibinfo{journal}{Phys. Rev.} \textbf{\bibinfo{volume}{D78}},
  \bibinfo{pages}{023502} (\bibinfo{year}{2008}), \eprint{0711.2906}.

\bibitem[{\citenamefont{Strassler and Zurek}(2007)}]{Strassler:2006im}
\bibinfo{author}{\bibfnamefont{M.~J.} \bibnamefont{Strassler}}
  \bibnamefont{and} \bibinfo{author}{\bibfnamefont{K.~M.} \bibnamefont{Zurek}},
  \bibinfo{journal}{Phys. Lett.} \textbf{\bibinfo{volume}{B651}},
  \bibinfo{pages}{374} (\bibinfo{year}{2007}), \eprint{hep-ph/0604261}.

\bibitem[{\citenamefont{Ma and Bertschinger}(1995)}]{Ma:1995ey}
\bibinfo{author}{\bibfnamefont{C.-P.} \bibnamefont{Ma}} \bibnamefont{and}
  \bibinfo{author}{\bibfnamefont{E.}~\bibnamefont{Bertschinger}},
  \bibinfo{journal}{Astrophys. J.} \textbf{\bibinfo{volume}{455}},
  \bibinfo{pages}{7} (\bibinfo{year}{1995}), \eprint{astro-ph/9506072}.

\bibitem[{\citenamefont{Kofman et~al.}(1986)\citenamefont{Kofman, Pogosyan, and
  Starobinsky}}]{Kofman:1986am}
\bibinfo{author}{\bibfnamefont{L.}~\bibnamefont{Kofman}},
  \bibinfo{author}{\bibfnamefont{D.}~\bibnamefont{Pogosyan}}, \bibnamefont{and}
  \bibinfo{author}{\bibfnamefont{A.~A.} \bibnamefont{Starobinsky}},
  \bibinfo{journal}{Sov. Astron. Lett.} \textbf{\bibinfo{volume}{12}},
  \bibinfo{pages}{175} (\bibinfo{year}{1986}).

\bibitem[{\citenamefont{{Bond} and {Efstathiou}}(1984)}]{Bond1984}
\bibinfo{author}{\bibfnamefont{J.~R.} \bibnamefont{{Bond}}} \bibnamefont{and}
  \bibinfo{author}{\bibfnamefont{G.}~\bibnamefont{{Efstathiou}}},
  \bibinfo{journal}{\apjl} \textbf{\bibinfo{volume}{285}}, \bibinfo{pages}{L45}
  (\bibinfo{year}{1984}).

\bibitem[{\citenamefont{Chen and Kamionkowski}(2004)}]{Chen:2003gz}
\bibinfo{author}{\bibfnamefont{X.-L.} \bibnamefont{Chen}} \bibnamefont{and}
  \bibinfo{author}{\bibfnamefont{M.}~\bibnamefont{Kamionkowski}},
  \bibinfo{journal}{Phys. Rev.} \textbf{\bibinfo{volume}{D70}},
  \bibinfo{pages}{043502} (\bibinfo{year}{2004}), \eprint{astro-ph/0310473}.

\bibitem[{\citenamefont{Schleicher et~al.}(2008)\citenamefont{Schleicher,
  Banerjee, and Klesser}}]{Schleicher:2008dt}
\bibinfo{author}{\bibfnamefont{D.~R.~G.} \bibnamefont{Schleicher}},
  \bibinfo{author}{\bibfnamefont{R.}~\bibnamefont{Banerjee}}, \bibnamefont{and}
  \bibinfo{author}{\bibfnamefont{R.~S.} \bibnamefont{Klesser}},
  \bibinfo{journal}{Phys. Rev.} \textbf{\bibinfo{volume}{D78}},
  \bibinfo{pages}{083005} (\bibinfo{year}{2008}), \eprint{0807.3802}.

\bibitem[{\citenamefont{{Seager} et~al.}(1999)\citenamefont{{Seager},
  {Sasselov}, and {Scott}}}]{Seager1999}
\bibinfo{author}{\bibfnamefont{S.}~\bibnamefont{{Seager}}},
  \bibinfo{author}{\bibfnamefont{D.~D.} \bibnamefont{{Sasselov}}},
  \bibnamefont{and} \bibinfo{author}{\bibfnamefont{D.}~\bibnamefont{{Scott}}},
  \bibinfo{journal}{\apjl} \textbf{\bibinfo{volume}{523}}, \bibinfo{pages}{L1}
  (\bibinfo{year}{1999}), \eprint{arXiv:astro-ph/9909275}.

\bibitem[{\citenamefont{{Seager} et~al.}(2000)\citenamefont{{Seager},
  {Sasselov}, and {Scott}}}]{Seager2000}
\bibinfo{author}{\bibfnamefont{S.}~\bibnamefont{{Seager}}},
  \bibinfo{author}{\bibfnamefont{D.~D.} \bibnamefont{{Sasselov}}},
  \bibnamefont{and} \bibinfo{author}{\bibfnamefont{D.}~\bibnamefont{{Scott}}},
  \bibinfo{journal}{\apjs} \textbf{\bibinfo{volume}{128}}, \bibinfo{pages}{407}
  (\bibinfo{year}{2000}), \eprint{arXiv:astro-ph/9912182}.

\bibitem[{\citenamefont{{Lewis} and {Bridle}}(2002)}]{Lewis2002}
\bibinfo{author}{\bibfnamefont{A.}~\bibnamefont{{Lewis}}} \bibnamefont{and}
  \bibinfo{author}{\bibfnamefont{S.}~\bibnamefont{{Bridle}}},
  \bibinfo{journal}{\prd} \textbf{\bibinfo{volume}{66}},
  \bibinfo{pages}{103511} (\bibinfo{year}{2002}),
  \eprint{arXiv:astro-ph/0205436}.

\bibitem[{\citenamefont{{Jones} et~al.}(2006)\citenamefont{{Jones}, {Ade},
  {Bock}, {Bond}, {Borrill}, {Boscaleri}, {Cabella}, {Contaldi}, {Crill}, {de
  Bernardis} et~al.}}]{Jones2006}
\bibinfo{author}{\bibfnamefont{W.~C.} \bibnamefont{{Jones}}},
  \bibinfo{author}{\bibfnamefont{P.~A.~R.} \bibnamefont{{Ade}}},
  \bibinfo{author}{\bibfnamefont{J.~J.} \bibnamefont{{Bock}}},
  \bibinfo{author}{\bibfnamefont{J.~R.} \bibnamefont{{Bond}}},
  \bibinfo{author}{\bibfnamefont{J.}~\bibnamefont{{Borrill}}},
  \bibinfo{author}{\bibfnamefont{A.}~\bibnamefont{{Boscaleri}}},
  \bibinfo{author}{\bibfnamefont{P.}~\bibnamefont{{Cabella}}},
  \bibinfo{author}{\bibfnamefont{C.~R.} \bibnamefont{{Contaldi}}},
  \bibinfo{author}{\bibfnamefont{B.~P.} \bibnamefont{{Crill}}},
  \bibinfo{author}{\bibfnamefont{P.}~\bibnamefont{{de Bernardis}}},
  \bibnamefont{et~al.}, \bibinfo{journal}{New Astronomy Review}
  \textbf{\bibinfo{volume}{50}}, \bibinfo{pages}{945} (\bibinfo{year}{2006}).

\bibitem[{\citenamefont{{Piacentini} et~al.}(2006)\citenamefont{{Piacentini},
  {Ade}, {Bock}, {Bond}, {Borrill}, {Boscaleri}, {Cabella}, {Contaldi},
  {Crill}, {de Bernardis} et~al.}}]{Piacentini2006}
\bibinfo{author}{\bibfnamefont{F.}~\bibnamefont{{Piacentini}}},
  \bibinfo{author}{\bibfnamefont{P.~A.~R.} \bibnamefont{{Ade}}},
  \bibinfo{author}{\bibfnamefont{J.~J.} \bibnamefont{{Bock}}},
  \bibinfo{author}{\bibfnamefont{J.~R.} \bibnamefont{{Bond}}},
  \bibinfo{author}{\bibfnamefont{J.}~\bibnamefont{{Borrill}}},
  \bibinfo{author}{\bibfnamefont{A.}~\bibnamefont{{Boscaleri}}},
  \bibinfo{author}{\bibfnamefont{P.}~\bibnamefont{{Cabella}}},
  \bibinfo{author}{\bibfnamefont{C.~R.} \bibnamefont{{Contaldi}}},
  \bibinfo{author}{\bibfnamefont{B.~P.} \bibnamefont{{Crill}}},
  \bibinfo{author}{\bibfnamefont{P.}~\bibnamefont{{de Bernardis}}},
  \bibnamefont{et~al.}, \bibinfo{journal}{\apj} \textbf{\bibinfo{volume}{647}},
  \bibinfo{pages}{833} (\bibinfo{year}{2006}), \eprint{arXiv:astro-ph/0507507}.

\bibitem[{\citenamefont{{Montroy} et~al.}(2006)\citenamefont{{Montroy}, {Ade},
  {Bock}, {Bond}, {Borrill}, {Boscaleri}, {Cabella}, {Contaldi}, {Crill}, {de
  Bernardis} et~al.}}]{Montroy2006}
\bibinfo{author}{\bibfnamefont{T.~E.} \bibnamefont{{Montroy}}},
  \bibinfo{author}{\bibfnamefont{P.~A.~R.} \bibnamefont{{Ade}}},
  \bibinfo{author}{\bibfnamefont{J.~J.} \bibnamefont{{Bock}}},
  \bibinfo{author}{\bibfnamefont{J.~R.} \bibnamefont{{Bond}}},
  \bibinfo{author}{\bibfnamefont{J.}~\bibnamefont{{Borrill}}},
  \bibinfo{author}{\bibfnamefont{A.}~\bibnamefont{{Boscaleri}}},
  \bibinfo{author}{\bibfnamefont{P.}~\bibnamefont{{Cabella}}},
  \bibinfo{author}{\bibfnamefont{C.~R.} \bibnamefont{{Contaldi}}},
  \bibinfo{author}{\bibfnamefont{B.~P.} \bibnamefont{{Crill}}},
  \bibinfo{author}{\bibfnamefont{P.}~\bibnamefont{{de Bernardis}}},
  \bibnamefont{et~al.}, \bibinfo{journal}{\apj} \textbf{\bibinfo{volume}{647}},
  \bibinfo{pages}{813} (\bibinfo{year}{2006}), \eprint{arXiv:astro-ph/0507514}.

\bibitem[{\citenamefont{Reichardt et~al.}(2008)}]{Reichardt2008}
\bibinfo{author}{\bibfnamefont{C.~L.} \bibnamefont{Reichardt}}
  \bibnamefont{et~al.} (\bibinfo{year}{2008}), \eprint{0801.1491}.

\bibitem[{\citenamefont{{Kuo} et~al.}(2006)\citenamefont{{Kuo}, {Ade}, {Bock},
  {Bond}, {Contaldi}, {Daub}, {Goldstein}, {Holzapfel}, {Lange}, {Lueker}
  et~al.}}]{Kuo2006}
\bibinfo{author}{\bibfnamefont{C.~L.} \bibnamefont{{Kuo}}},
  \bibinfo{author}{\bibfnamefont{P.~A.~R.} \bibnamefont{{Ade}}},
  \bibinfo{author}{\bibfnamefont{J.~J.} \bibnamefont{{Bock}}},
  \bibinfo{author}{\bibfnamefont{J.~R.} \bibnamefont{{Bond}}},
  \bibinfo{author}{\bibfnamefont{C.~R.} \bibnamefont{{Contaldi}}},
  \bibinfo{author}{\bibfnamefont{M.~D.} \bibnamefont{{Daub}}},
  \bibinfo{author}{\bibfnamefont{J.~H.} \bibnamefont{{Goldstein}}},
  \bibinfo{author}{\bibfnamefont{W.~L.} \bibnamefont{{Holzapfel}}},
  \bibinfo{author}{\bibfnamefont{A.~E.} \bibnamefont{{Lange}}},
  \bibinfo{author}{\bibfnamefont{M.}~\bibnamefont{{Lueker}}},
  \bibnamefont{et~al.} (\bibinfo{year}{2006}), \eprint{astro-ph/0611198}.

\bibitem[{\citenamefont{{Runyan} et~al.}(2003)\citenamefont{{Runyan}, {Ade},
  {Bhatia}, {Bock}, {Daub}, {Goldstein}, {Haynes}, {Holzapfel}, {Kuo}, {Lange}
  et~al.}}]{Runyan2003}
\bibinfo{author}{\bibfnamefont{M.~C.} \bibnamefont{{Runyan}}},
  \bibinfo{author}{\bibfnamefont{P.~A.~R.} \bibnamefont{{Ade}}},
  \bibinfo{author}{\bibfnamefont{R.~S.} \bibnamefont{{Bhatia}}},
  \bibinfo{author}{\bibfnamefont{J.~J.} \bibnamefont{{Bock}}},
  \bibinfo{author}{\bibfnamefont{M.~D.} \bibnamefont{{Daub}}},
  \bibinfo{author}{\bibfnamefont{J.~H.} \bibnamefont{{Goldstein}}},
  \bibinfo{author}{\bibfnamefont{C.~V.} \bibnamefont{{Haynes}}},
  \bibinfo{author}{\bibfnamefont{W.~L.} \bibnamefont{{Holzapfel}}},
  \bibinfo{author}{\bibfnamefont{C.~L.} \bibnamefont{{Kuo}}},
  \bibinfo{author}{\bibfnamefont{A.~E.} \bibnamefont{{Lange}}},
  \bibnamefont{et~al.}, \bibinfo{journal}{\apjs}
  \textbf{\bibinfo{volume}{149}}, \bibinfo{pages}{265} (\bibinfo{year}{2003}),
  \eprint{arXiv:astro-ph/0303515}.

\bibitem[{\citenamefont{{Goldstein} et~al.}(2003)\citenamefont{{Goldstein},
  {Ade}, {Bock}, {Bond}, {Cantalupo}, {Contaldi}, {Daub}, {Holzapfel}, {Kuo},
  {Lange} et~al.}}]{Goldstein2003}
\bibinfo{author}{\bibfnamefont{J.~H.} \bibnamefont{{Goldstein}}},
  \bibinfo{author}{\bibfnamefont{P.~A.~R.} \bibnamefont{{Ade}}},
  \bibinfo{author}{\bibfnamefont{J.~J.} \bibnamefont{{Bock}}},
  \bibinfo{author}{\bibfnamefont{J.~R.} \bibnamefont{{Bond}}},
  \bibinfo{author}{\bibfnamefont{C.}~\bibnamefont{{Cantalupo}}},
  \bibinfo{author}{\bibfnamefont{C.~R.} \bibnamefont{{Contaldi}}},
  \bibinfo{author}{\bibfnamefont{M.~D.} \bibnamefont{{Daub}}},
  \bibinfo{author}{\bibfnamefont{W.~L.} \bibnamefont{{Holzapfel}}},
  \bibinfo{author}{\bibfnamefont{C.}~\bibnamefont{{Kuo}}},
  \bibinfo{author}{\bibfnamefont{A.~E.} \bibnamefont{{Lange}}},
  \bibnamefont{et~al.}, \bibinfo{journal}{\apj} \textbf{\bibinfo{volume}{599}},
  \bibinfo{pages}{773} (\bibinfo{year}{2003}), \eprint{arXiv:astro-ph/0212517}.

\bibitem[{\citenamefont{{Pearson} et~al.}(2003)\citenamefont{{Pearson},
  {Mason}, {Readhead}, {Shepherd}, {Sievers}, {Udomprasert}, {Cartwright},
  {Farmer}, {Padin}, {Myers} et~al.}}]{Pearson2003}
\bibinfo{author}{\bibfnamefont{T.~J.} \bibnamefont{{Pearson}}},
  \bibinfo{author}{\bibfnamefont{B.~S.} \bibnamefont{{Mason}}},
  \bibinfo{author}{\bibfnamefont{A.~C.~S.} \bibnamefont{{Readhead}}},
  \bibinfo{author}{\bibfnamefont{M.~C.} \bibnamefont{{Shepherd}}},
  \bibinfo{author}{\bibfnamefont{J.~L.} \bibnamefont{{Sievers}}},
  \bibinfo{author}{\bibfnamefont{P.~S.} \bibnamefont{{Udomprasert}}},
  \bibinfo{author}{\bibfnamefont{J.~K.} \bibnamefont{{Cartwright}}},
  \bibinfo{author}{\bibfnamefont{A.~J.} \bibnamefont{{Farmer}}},
  \bibinfo{author}{\bibfnamefont{S.}~\bibnamefont{{Padin}}},
  \bibinfo{author}{\bibfnamefont{S.~T.} \bibnamefont{{Myers}}},
  \bibnamefont{et~al.}, \bibinfo{journal}{\apj} \textbf{\bibinfo{volume}{591}},
  \bibinfo{pages}{556} (\bibinfo{year}{2003}), \eprint{arXiv:astro-ph/0205388}.

\bibitem[{\citenamefont{{Readhead}
  et~al.}(2004{\natexlab{a}})\citenamefont{{Readhead}, {Mason}, {Contaldi},
  {Pearson}, {Bond}, {Myers}, {Padin}, {Sievers}, {Cartwright}, {Shepherd}
  et~al.}}]{Readhead2004a}
\bibinfo{author}{\bibfnamefont{A.~C.~S.} \bibnamefont{{Readhead}}},
  \bibinfo{author}{\bibfnamefont{B.~S.} \bibnamefont{{Mason}}},
  \bibinfo{author}{\bibfnamefont{C.~R.} \bibnamefont{{Contaldi}}},
  \bibinfo{author}{\bibfnamefont{T.~J.} \bibnamefont{{Pearson}}},
  \bibinfo{author}{\bibfnamefont{J.~R.} \bibnamefont{{Bond}}},
  \bibinfo{author}{\bibfnamefont{S.~T.} \bibnamefont{{Myers}}},
  \bibinfo{author}{\bibfnamefont{S.}~\bibnamefont{{Padin}}},
  \bibinfo{author}{\bibfnamefont{J.~L.} \bibnamefont{{Sievers}}},
  \bibinfo{author}{\bibfnamefont{J.~K.} \bibnamefont{{Cartwright}}},
  \bibinfo{author}{\bibfnamefont{M.~C.} \bibnamefont{{Shepherd}}},
  \bibnamefont{et~al.}, \bibinfo{journal}{\apj} \textbf{\bibinfo{volume}{609}},
  \bibinfo{pages}{498} (\bibinfo{year}{2004}{\natexlab{a}}),
  \eprint{arXiv:astro-ph/0402359}.

\bibitem[{\citenamefont{{Readhead}
  et~al.}(2004{\natexlab{b}})\citenamefont{{Readhead}, {Myers}, {Pearson},
  {Sievers}, {Mason}, {Contaldi}, {Bond}, {Bustos}, {Altamirano}, {Achermann}
  et~al.}}]{Readhead2004b}
\bibinfo{author}{\bibfnamefont{A.~C.~S.} \bibnamefont{{Readhead}}},
  \bibinfo{author}{\bibfnamefont{S.~T.} \bibnamefont{{Myers}}},
  \bibinfo{author}{\bibfnamefont{T.~J.} \bibnamefont{{Pearson}}},
  \bibinfo{author}{\bibfnamefont{J.~L.} \bibnamefont{{Sievers}}},
  \bibinfo{author}{\bibfnamefont{B.~S.} \bibnamefont{{Mason}}},
  \bibinfo{author}{\bibfnamefont{C.~R.} \bibnamefont{{Contaldi}}},
  \bibinfo{author}{\bibfnamefont{J.~R.} \bibnamefont{{Bond}}},
  \bibinfo{author}{\bibfnamefont{R.}~\bibnamefont{{Bustos}}},
  \bibinfo{author}{\bibfnamefont{P.}~\bibnamefont{{Altamirano}}},
  \bibinfo{author}{\bibfnamefont{C.}~\bibnamefont{{Achermann}}},
  \bibnamefont{et~al.}, \bibinfo{journal}{Science}
  \textbf{\bibinfo{volume}{306}}, \bibinfo{pages}{836}
  (\bibinfo{year}{2004}{\natexlab{b}}), \eprint{arXiv:astro-ph/0409569}.

\bibitem[{\citenamefont{{Sievers} et~al.}(2007)\citenamefont{{Sievers},
  {Achermann}, {Bond}, {Bronfman}, {Bustos}, {Contaldi}, {Dickinson},
  {Ferreira}, {Jones}, {Lewis} et~al.}}]{Sievers2007}
\bibinfo{author}{\bibfnamefont{J.~L.} \bibnamefont{{Sievers}}},
  \bibinfo{author}{\bibfnamefont{C.}~\bibnamefont{{Achermann}}},
  \bibinfo{author}{\bibfnamefont{J.~R.} \bibnamefont{{Bond}}},
  \bibinfo{author}{\bibfnamefont{L.}~\bibnamefont{{Bronfman}}},
  \bibinfo{author}{\bibfnamefont{R.}~\bibnamefont{{Bustos}}},
  \bibinfo{author}{\bibfnamefont{C.~R.} \bibnamefont{{Contaldi}}},
  \bibinfo{author}{\bibfnamefont{C.}~\bibnamefont{{Dickinson}}},
  \bibinfo{author}{\bibfnamefont{P.~G.} \bibnamefont{{Ferreira}}},
  \bibinfo{author}{\bibfnamefont{M.~E.} \bibnamefont{{Jones}}},
  \bibinfo{author}{\bibfnamefont{A.~M.} \bibnamefont{{Lewis}}},
  \bibnamefont{et~al.}, \bibinfo{journal}{\apj} \textbf{\bibinfo{volume}{660}},
  \bibinfo{pages}{976} (\bibinfo{year}{2007}).

\bibitem[{\citenamefont{{Dickinson} et~al.}(2004)\citenamefont{{Dickinson},
  {Battye}, {Carreira}, {Cleary}, {Davies}, {Davis}, {Genova-Santos},
  {Grainge}, {Guti{\'e}rrez}, {Hafez} et~al.}}]{Dickinson2004}
\bibinfo{author}{\bibfnamefont{C.}~\bibnamefont{{Dickinson}}},
  \bibinfo{author}{\bibfnamefont{R.~A.} \bibnamefont{{Battye}}},
  \bibinfo{author}{\bibfnamefont{P.}~\bibnamefont{{Carreira}}},
  \bibinfo{author}{\bibfnamefont{K.}~\bibnamefont{{Cleary}}},
  \bibinfo{author}{\bibfnamefont{R.~D.} \bibnamefont{{Davies}}},
  \bibinfo{author}{\bibfnamefont{R.~J.} \bibnamefont{{Davis}}},
  \bibinfo{author}{\bibfnamefont{R.}~\bibnamefont{{Genova-Santos}}},
  \bibinfo{author}{\bibfnamefont{K.}~\bibnamefont{{Grainge}}},
  \bibinfo{author}{\bibfnamefont{C.~M.} \bibnamefont{{Guti{\'e}rrez}}},
  \bibinfo{author}{\bibfnamefont{Y.~A.} \bibnamefont{{Hafez}}},
  \bibnamefont{et~al.}, \bibinfo{journal}{\mnras}
  \textbf{\bibinfo{volume}{353}}, \bibinfo{pages}{732} (\bibinfo{year}{2004}),
  \eprint{arXiv:astro-ph/0402498}.

\bibitem[{\citenamefont{{Halverson} et~al.}(2002)\citenamefont{{Halverson},
  {Leitch}, {Pryke}, {Kovac}, {Carlstrom}, {Holzapfel}, {Dragovan},
  {Cartwright}, {Mason}, {Padin} et~al.}}]{Halverson2002}
\bibinfo{author}{\bibfnamefont{N.~W.} \bibnamefont{{Halverson}}},
  \bibinfo{author}{\bibfnamefont{E.~M.} \bibnamefont{{Leitch}}},
  \bibinfo{author}{\bibfnamefont{C.}~\bibnamefont{{Pryke}}},
  \bibinfo{author}{\bibfnamefont{J.}~\bibnamefont{{Kovac}}},
  \bibinfo{author}{\bibfnamefont{J.~E.} \bibnamefont{{Carlstrom}}},
  \bibinfo{author}{\bibfnamefont{W.~L.} \bibnamefont{{Holzapfel}}},
  \bibinfo{author}{\bibfnamefont{M.}~\bibnamefont{{Dragovan}}},
  \bibinfo{author}{\bibfnamefont{J.~K.} \bibnamefont{{Cartwright}}},
  \bibinfo{author}{\bibfnamefont{B.~S.} \bibnamefont{{Mason}}},
  \bibinfo{author}{\bibfnamefont{S.}~\bibnamefont{{Padin}}},
  \bibnamefont{et~al.}, \bibinfo{journal}{\apj} \textbf{\bibinfo{volume}{568}},
  \bibinfo{pages}{38} (\bibinfo{year}{2002}), \eprint{arXiv:astro-ph/0104489}.

\bibitem[{\citenamefont{{Leitch} et~al.}(2005)\citenamefont{{Leitch}, {Kovac},
  {Halverson}, {Carlstrom}, {Pryke}, and {Smith}}}]{Leitch2005}
\bibinfo{author}{\bibfnamefont{E.~M.} \bibnamefont{{Leitch}}},
  \bibinfo{author}{\bibfnamefont{J.~M.} \bibnamefont{{Kovac}}},
  \bibinfo{author}{\bibfnamefont{N.~W.} \bibnamefont{{Halverson}}},
  \bibinfo{author}{\bibfnamefont{J.~E.} \bibnamefont{{Carlstrom}}},
  \bibinfo{author}{\bibfnamefont{C.}~\bibnamefont{{Pryke}}}, \bibnamefont{and}
  \bibinfo{author}{\bibfnamefont{M.~W.~E.} \bibnamefont{{Smith}}},
  \bibinfo{journal}{\apj} \textbf{\bibinfo{volume}{624}}, \bibinfo{pages}{10}
  (\bibinfo{year}{2005}), \eprint{arXiv:astro-ph/0409357}.

\bibitem[{\citenamefont{{Hanany} et~al.}(2000)\citenamefont{{Hanany}, {Ade},
  {Balbi}, {Bock}, {Borrill}, {Boscaleri}, {de Bernardis}, {Ferreira},
  {Hristov}, {Jaffe} et~al.}}]{Hanany2000}
\bibinfo{author}{\bibfnamefont{S.}~\bibnamefont{{Hanany}}},
  \bibinfo{author}{\bibfnamefont{P.}~\bibnamefont{{Ade}}},
  \bibinfo{author}{\bibfnamefont{A.}~\bibnamefont{{Balbi}}},
  \bibinfo{author}{\bibfnamefont{J.}~\bibnamefont{{Bock}}},
  \bibinfo{author}{\bibfnamefont{J.}~\bibnamefont{{Borrill}}},
  \bibinfo{author}{\bibfnamefont{A.}~\bibnamefont{{Boscaleri}}},
  \bibinfo{author}{\bibfnamefont{P.}~\bibnamefont{{de Bernardis}}},
  \bibinfo{author}{\bibfnamefont{P.~G.} \bibnamefont{{Ferreira}}},
  \bibinfo{author}{\bibfnamefont{V.~V.} \bibnamefont{{Hristov}}},
  \bibinfo{author}{\bibfnamefont{A.~H.} \bibnamefont{{Jaffe}}},
  \bibnamefont{et~al.}, \bibinfo{journal}{\apjl}
  \textbf{\bibinfo{volume}{545}}, \bibinfo{pages}{L5} (\bibinfo{year}{2000}),
  \eprint{arXiv:astro-ph/0005123}.

\bibitem[{\citenamefont{{Bond} et~al.}(2005)\citenamefont{{Bond}, {Contaldi},
  {Pen}, {Pogosyan}, {Prunet}, {Ruetalo}, {Wadsley}, {Zhang}, {Mason}, {Myers}
  et~al.}}]{Bond2005}
\bibinfo{author}{\bibfnamefont{J.~R.} \bibnamefont{{Bond}}},
  \bibinfo{author}{\bibfnamefont{C.~R.} \bibnamefont{{Contaldi}}},
  \bibinfo{author}{\bibfnamefont{U.-L.} \bibnamefont{{Pen}}},
  \bibinfo{author}{\bibfnamefont{D.}~\bibnamefont{{Pogosyan}}},
  \bibinfo{author}{\bibfnamefont{S.}~\bibnamefont{{Prunet}}},
  \bibinfo{author}{\bibfnamefont{M.~I.} \bibnamefont{{Ruetalo}}},
  \bibinfo{author}{\bibfnamefont{J.~W.} \bibnamefont{{Wadsley}}},
  \bibinfo{author}{\bibfnamefont{P.}~\bibnamefont{{Zhang}}},
  \bibinfo{author}{\bibfnamefont{B.~S.} \bibnamefont{{Mason}}},
  \bibinfo{author}{\bibfnamefont{S.~T.} \bibnamefont{{Myers}}},
  \bibnamefont{et~al.}, \bibinfo{journal}{\apj} \textbf{\bibinfo{volume}{626}},
  \bibinfo{pages}{12} (\bibinfo{year}{2005}), \eprint{arXiv:astro-ph/0205386}.

\bibitem[{\citenamefont{Kowalski et~al.}(2008)}]{Kowalski2008}
\bibinfo{author}{\bibfnamefont{M.}~\bibnamefont{Kowalski}} \bibnamefont{et~al.}
  (\bibinfo{year}{2008}), \eprint{0804.4142}.

\bibitem[{\citenamefont{{Cole} et~al.}(2005)\citenamefont{{Cole}, {Percival},
  {Peacock}, {Norberg}, {Baugh}, {Frenk}, {Baldry}, {Bland-Hawthorn},
  {Bridges}, {Cannon} et~al.}}]{Cole2005}
\bibinfo{author}{\bibfnamefont{S.}~\bibnamefont{{Cole}}},
  \bibinfo{author}{\bibfnamefont{W.~J.} \bibnamefont{{Percival}}},
  \bibinfo{author}{\bibfnamefont{J.~A.} \bibnamefont{{Peacock}}},
  \bibinfo{author}{\bibfnamefont{P.}~\bibnamefont{{Norberg}}},
  \bibinfo{author}{\bibfnamefont{C.~M.} \bibnamefont{{Baugh}}},
  \bibinfo{author}{\bibfnamefont{C.~S.} \bibnamefont{{Frenk}}},
  \bibinfo{author}{\bibfnamefont{I.}~\bibnamefont{{Baldry}}},
  \bibinfo{author}{\bibfnamefont{J.}~\bibnamefont{{Bland-Hawthorn}}},
  \bibinfo{author}{\bibfnamefont{T.}~\bibnamefont{{Bridges}}},
  \bibinfo{author}{\bibfnamefont{R.}~\bibnamefont{{Cannon}}},
  \bibnamefont{et~al.}, \bibinfo{journal}{\mnras}
  \textbf{\bibinfo{volume}{362}}, \bibinfo{pages}{505} (\bibinfo{year}{2005}),
  \eprint{arXiv:astro-ph/0501174}.

\bibitem[{\citenamefont{{Tegmark} et~al.}(2006)\citenamefont{{Tegmark},
  {Eisenstein}, {Strauss}, {Weinberg}, {Blanton}, {Frieman}, {Fukugita},
  {Gunn}, {Hamilton}, {Knapp} et~al.}}]{Tegmark2006}
\bibinfo{author}{\bibfnamefont{M.}~\bibnamefont{{Tegmark}}},
  \bibinfo{author}{\bibfnamefont{D.~J.} \bibnamefont{{Eisenstein}}},
  \bibinfo{author}{\bibfnamefont{M.~A.} \bibnamefont{{Strauss}}},
  \bibinfo{author}{\bibfnamefont{D.~H.} \bibnamefont{{Weinberg}}},
  \bibinfo{author}{\bibfnamefont{M.~R.} \bibnamefont{{Blanton}}},
  \bibinfo{author}{\bibfnamefont{J.~A.} \bibnamefont{{Frieman}}},
  \bibinfo{author}{\bibfnamefont{M.}~\bibnamefont{{Fukugita}}},
  \bibinfo{author}{\bibfnamefont{J.~E.} \bibnamefont{{Gunn}}},
  \bibinfo{author}{\bibfnamefont{A.~J.~S.} \bibnamefont{{Hamilton}}},
  \bibinfo{author}{\bibfnamefont{G.~R.} \bibnamefont{{Knapp}}},
  \bibnamefont{et~al.}, \bibinfo{journal}{\prd} \textbf{\bibinfo{volume}{74}},
  \bibinfo{pages}{123507} (\bibinfo{year}{2006}),
  \eprint{arXiv:astro-ph/0608632}.

\bibitem[{\citenamefont{{Eisenstein} et~al.}(2005)\citenamefont{{Eisenstein},
  {Zehavi}, {Hogg}, {Scoccimarro}, {Blanton}, {Nichol}, {Scranton}, {Seo},
  {Tegmark}, {Zheng} et~al.}}]{Eisenstein2005}
\bibinfo{author}{\bibfnamefont{D.~J.} \bibnamefont{{Eisenstein}}},
  \bibinfo{author}{\bibfnamefont{I.}~\bibnamefont{{Zehavi}}},
  \bibinfo{author}{\bibfnamefont{D.~W.} \bibnamefont{{Hogg}}},
  \bibinfo{author}{\bibfnamefont{R.}~\bibnamefont{{Scoccimarro}}},
  \bibinfo{author}{\bibfnamefont{M.~R.} \bibnamefont{{Blanton}}},
  \bibinfo{author}{\bibfnamefont{R.~C.} \bibnamefont{{Nichol}}},
  \bibinfo{author}{\bibfnamefont{R.}~\bibnamefont{{Scranton}}},
  \bibinfo{author}{\bibfnamefont{H.-J.} \bibnamefont{{Seo}}},
  \bibinfo{author}{\bibfnamefont{M.}~\bibnamefont{{Tegmark}}},
  \bibinfo{author}{\bibfnamefont{Z.}~\bibnamefont{{Zheng}}},
  \bibnamefont{et~al.}, \bibinfo{journal}{\apj} \textbf{\bibinfo{volume}{633}},
  \bibinfo{pages}{560} (\bibinfo{year}{2005}), \eprint{arXiv:astro-ph/0501171}.

\bibitem[{\citenamefont{{Percival} et~al.}(2007)\citenamefont{{Percival},
  {Nichol}, {Eisenstein}, {Frieman}, {Fukugita}, {Loveday}, {Pope},
  {Schneider}, {Szalay}, {Tegmark} et~al.}}]{Percival2007}
\bibinfo{author}{\bibfnamefont{W.~J.} \bibnamefont{{Percival}}},
  \bibinfo{author}{\bibfnamefont{R.~C.} \bibnamefont{{Nichol}}},
  \bibinfo{author}{\bibfnamefont{D.~J.} \bibnamefont{{Eisenstein}}},
  \bibinfo{author}{\bibfnamefont{J.~A.} \bibnamefont{{Frieman}}},
  \bibinfo{author}{\bibfnamefont{M.}~\bibnamefont{{Fukugita}}},
  \bibinfo{author}{\bibfnamefont{J.}~\bibnamefont{{Loveday}}},
  \bibinfo{author}{\bibfnamefont{A.~C.} \bibnamefont{{Pope}}},
  \bibinfo{author}{\bibfnamefont{D.~P.} \bibnamefont{{Schneider}}},
  \bibinfo{author}{\bibfnamefont{A.~S.} \bibnamefont{{Szalay}}},
  \bibinfo{author}{\bibfnamefont{M.}~\bibnamefont{{Tegmark}}},
  \bibnamefont{et~al.}, \bibinfo{journal}{\apj} \textbf{\bibinfo{volume}{657}},
  \bibinfo{pages}{645} (\bibinfo{year}{2007}), \eprint{arXiv:astro-ph/0608636}.

\bibitem[{\citenamefont{{Massey} et~al.}(2007)\citenamefont{{Massey}, {Rhodes},
  {Leauthaud}, {Capak}, {Ellis}, {Koekemoer}, {R{\'e}fr{\'e}gier}, {Scoville},
  {Taylor}, {Albert} et~al.}}]{Massey2007}
\bibinfo{author}{\bibfnamefont{R.}~\bibnamefont{{Massey}}},
  \bibinfo{author}{\bibfnamefont{J.}~\bibnamefont{{Rhodes}}},
  \bibinfo{author}{\bibfnamefont{A.}~\bibnamefont{{Leauthaud}}},
  \bibinfo{author}{\bibfnamefont{P.}~\bibnamefont{{Capak}}},
  \bibinfo{author}{\bibfnamefont{R.}~\bibnamefont{{Ellis}}},
  \bibinfo{author}{\bibfnamefont{A.}~\bibnamefont{{Koekemoer}}},
  \bibinfo{author}{\bibfnamefont{A.}~\bibnamefont{{R{\'e}fr{\'e}gier}}},
  \bibinfo{author}{\bibfnamefont{N.}~\bibnamefont{{Scoville}}},
  \bibinfo{author}{\bibfnamefont{J.~E.} \bibnamefont{{Taylor}}},
  \bibinfo{author}{\bibfnamefont{J.}~\bibnamefont{{Albert}}},
  \bibnamefont{et~al.}, \bibinfo{journal}{\apjs}
  \textbf{\bibinfo{volume}{172}}, \bibinfo{pages}{239} (\bibinfo{year}{2007}),
  \eprint{arXiv:astro-ph/0701480}.

\bibitem[{\citenamefont{{Lesgourgues} et~al.}(2007)\citenamefont{{Lesgourgues},
  {Viel}, {Haehnelt}, and {Massey}}}]{Lesgourgues2007}
\bibinfo{author}{\bibfnamefont{J.}~\bibnamefont{{Lesgourgues}}},
  \bibinfo{author}{\bibfnamefont{M.}~\bibnamefont{{Viel}}},
  \bibinfo{author}{\bibfnamefont{M.~G.} \bibnamefont{{Haehnelt}}},
  \bibnamefont{and} \bibinfo{author}{\bibfnamefont{R.}~\bibnamefont{{Massey}}},
  \bibinfo{journal}{Journal of Cosmology and Astro-Particle Physics}
  \textbf{\bibinfo{volume}{11}}, \bibinfo{pages}{8} (\bibinfo{year}{2007}),
  \eprint{arXiv:0705.0533}.

\bibitem[{\citenamefont{{Hoekstra} et~al.}(2006)\citenamefont{{Hoekstra},
  {Mellier}, {van Waerbeke}, {Semboloni}, {Fu}, {Hudson}, {Parker}, {Tereno},
  and {Benabed}}}]{Hoekstra2006}
\bibinfo{author}{\bibfnamefont{H.}~\bibnamefont{{Hoekstra}}},
  \bibinfo{author}{\bibfnamefont{Y.}~\bibnamefont{{Mellier}}},
  \bibinfo{author}{\bibfnamefont{L.}~\bibnamefont{{van Waerbeke}}},
  \bibinfo{author}{\bibfnamefont{E.}~\bibnamefont{{Semboloni}}},
  \bibinfo{author}{\bibfnamefont{L.}~\bibnamefont{{Fu}}},
  \bibinfo{author}{\bibfnamefont{M.~J.} \bibnamefont{{Hudson}}},
  \bibinfo{author}{\bibfnamefont{L.~C.} \bibnamefont{{Parker}}},
  \bibinfo{author}{\bibfnamefont{I.}~\bibnamefont{{Tereno}}}, \bibnamefont{and}
  \bibinfo{author}{\bibfnamefont{K.}~\bibnamefont{{Benabed}}},
  \bibinfo{journal}{\apj} \textbf{\bibinfo{volume}{647}}, \bibinfo{pages}{116}
  (\bibinfo{year}{2006}), \eprint{arXiv:astro-ph/0511089}.

\bibitem[{\citenamefont{Benjamin et~al.}(2007)}]{Benjamin2007}
\bibinfo{author}{\bibfnamefont{J.}~\bibnamefont{Benjamin}}
  \bibnamefont{et~al.}, \bibinfo{journal}{arXiv: astro-ph/0703570}
  (\bibinfo{year}{2007}), \eprint{astro-ph/0703570}.

\bibitem[{\citenamefont{{Schimd} et~al.}(2007)\citenamefont{{Schimd}, {Tereno},
  {Uzan}, {Mellier}, {van Waerbeke}, {Semboloni}, {Hoekstra}, {Fu}, and
  {Riazuelo}}}]{Schimd2007}
\bibinfo{author}{\bibfnamefont{C.}~\bibnamefont{{Schimd}}},
  \bibinfo{author}{\bibfnamefont{I.}~\bibnamefont{{Tereno}}},
  \bibinfo{author}{\bibfnamefont{J.-P.} \bibnamefont{{Uzan}}},
  \bibinfo{author}{\bibfnamefont{Y.}~\bibnamefont{{Mellier}}},
  \bibinfo{author}{\bibfnamefont{L.}~\bibnamefont{{van Waerbeke}}},
  \bibinfo{author}{\bibfnamefont{E.}~\bibnamefont{{Semboloni}}},
  \bibinfo{author}{\bibfnamefont{H.}~\bibnamefont{{Hoekstra}}},
  \bibinfo{author}{\bibfnamefont{L.}~\bibnamefont{{Fu}}}, \bibnamefont{and}
  \bibinfo{author}{\bibfnamefont{A.}~\bibnamefont{{Riazuelo}}},
  \bibinfo{journal}{\aap} \textbf{\bibinfo{volume}{463}}, \bibinfo{pages}{405}
  (\bibinfo{year}{2007}), \eprint{arXiv:astro-ph/0603158}.

\bibitem[{\citenamefont{{Hetterscheidt}
  et~al.}(2006)\citenamefont{{Hetterscheidt}, {Simon}, {Erben}, {Schneider},
  {Schirmer}, {Dietrich}, {Hildebrandt}, {Cordes}, {Schrabback}, {Haberzettl}
  et~al.}}]{Hetterscheidt2006}
\bibinfo{author}{\bibfnamefont{M.}~\bibnamefont{{Hetterscheidt}}},
  \bibinfo{author}{\bibfnamefont{P.}~\bibnamefont{{Simon}}},
  \bibinfo{author}{\bibfnamefont{T.}~\bibnamefont{{Erben}}},
  \bibinfo{author}{\bibfnamefont{P.}~\bibnamefont{{Schneider}}},
  \bibinfo{author}{\bibfnamefont{M.}~\bibnamefont{{Schirmer}}},
  \bibinfo{author}{\bibfnamefont{J.~P.} \bibnamefont{{Dietrich}}},
  \bibinfo{author}{\bibfnamefont{H.}~\bibnamefont{{Hildebrandt}}},
  \bibinfo{author}{\bibfnamefont{O.}~\bibnamefont{{Cordes}}},
  \bibinfo{author}{\bibfnamefont{T.}~\bibnamefont{{Schrabback}}},
  \bibinfo{author}{\bibfnamefont{L.}~\bibnamefont{{Haberzettl}}},
  \bibnamefont{et~al.}, \bibinfo{journal}{The Messenger}
  \textbf{\bibinfo{volume}{126}}, \bibinfo{pages}{19} (\bibinfo{year}{2006}).

\bibitem[{\citenamefont{{Hetterscheidt}
  et~al.}(2007)\citenamefont{{Hetterscheidt}, {Simon}, {Schirmer},
  {Hildebrandt}, {Schrabback}, {Erben}, and {Schneider}}}]{Hetterscheidt2007}
\bibinfo{author}{\bibfnamefont{M.}~\bibnamefont{{Hetterscheidt}}},
  \bibinfo{author}{\bibfnamefont{P.}~\bibnamefont{{Simon}}},
  \bibinfo{author}{\bibfnamefont{M.}~\bibnamefont{{Schirmer}}},
  \bibinfo{author}{\bibfnamefont{H.}~\bibnamefont{{Hildebrandt}}},
  \bibinfo{author}{\bibfnamefont{T.}~\bibnamefont{{Schrabback}}},
  \bibinfo{author}{\bibfnamefont{T.}~\bibnamefont{{Erben}}}, \bibnamefont{and}
  \bibinfo{author}{\bibfnamefont{P.}~\bibnamefont{{Schneider}}},
  \bibinfo{journal}{\aap} \textbf{\bibinfo{volume}{468}}, \bibinfo{pages}{859}
  (\bibinfo{year}{2007}), \eprint{arXiv:astro-ph/0606571}.

\bibitem[{\citenamefont{{Hoekstra}
  et~al.}(2002{\natexlab{a}})\citenamefont{{Hoekstra}, {Yee}, {Gladders},
  {Barrientos}, {Hall}, and {Infante}}}]{Hoekstra2002a}
\bibinfo{author}{\bibfnamefont{H.}~\bibnamefont{{Hoekstra}}},
  \bibinfo{author}{\bibfnamefont{H.~K.~C.} \bibnamefont{{Yee}}},
  \bibinfo{author}{\bibfnamefont{M.~D.} \bibnamefont{{Gladders}}},
  \bibinfo{author}{\bibfnamefont{L.~F.} \bibnamefont{{Barrientos}}},
  \bibinfo{author}{\bibfnamefont{P.~B.} \bibnamefont{{Hall}}},
  \bibnamefont{and}
  \bibinfo{author}{\bibfnamefont{L.}~\bibnamefont{{Infante}}},
  \bibinfo{journal}{\apj} \textbf{\bibinfo{volume}{572}}, \bibinfo{pages}{55}
  (\bibinfo{year}{2002}{\natexlab{a}}), \eprint{arXiv:astro-ph/0202285}.

\bibitem[{\citenamefont{{Hoekstra}
  et~al.}(2002{\natexlab{b}})\citenamefont{{Hoekstra}, {Yee}, and
  {Gladders}}}]{Hoekstra2002b}
\bibinfo{author}{\bibfnamefont{H.}~\bibnamefont{{Hoekstra}}},
  \bibinfo{author}{\bibfnamefont{H.~K.~C.} \bibnamefont{{Yee}}},
  \bibnamefont{and} \bibinfo{author}{\bibfnamefont{M.~D.}
  \bibnamefont{{Gladders}}}, \bibinfo{journal}{\apj}
  \textbf{\bibinfo{volume}{577}}, \bibinfo{pages}{595}
  (\bibinfo{year}{2002}{\natexlab{b}}), \eprint{arXiv:astro-ph/0204295}.

\bibitem[{\citenamefont{{Van Waerbeke} et~al.}(2005)\citenamefont{{Van
  Waerbeke}, {Mellier}, and {Hoekstra}}}]{Van-Waerbeke2005}
\bibinfo{author}{\bibfnamefont{L.}~\bibnamefont{{Van Waerbeke}}},
  \bibinfo{author}{\bibfnamefont{Y.}~\bibnamefont{{Mellier}}},
  \bibnamefont{and}
  \bibinfo{author}{\bibfnamefont{H.}~\bibnamefont{{Hoekstra}}},
  \bibinfo{journal}{\aap} \textbf{\bibinfo{volume}{429}}, \bibinfo{pages}{75}
  (\bibinfo{year}{2005}), \eprint{arXiv:astro-ph/0406468}.

\bibitem[{\citenamefont{{Kaiser}}(1992)}]{Kaiser1992}
\bibinfo{author}{\bibfnamefont{N.}~\bibnamefont{{Kaiser}}},
  \bibinfo{journal}{\apj} \textbf{\bibinfo{volume}{388}}, \bibinfo{pages}{272}
  (\bibinfo{year}{1992}).

\bibitem[{\citenamefont{{Kaiser}}(1998)}]{Kaiser1998}
\bibinfo{author}{\bibfnamefont{N.}~\bibnamefont{{Kaiser}}},
  \bibinfo{journal}{\apj} \textbf{\bibinfo{volume}{498}}, \bibinfo{pages}{26}
  (\bibinfo{year}{1998}), \eprint{arXiv:astro-ph/9610120}.

\bibitem[{\citenamefont{{Viel} et~al.}(2004)\citenamefont{{Viel}, {Haehnelt},
  and {Springel}}}]{Viel2004}
\bibinfo{author}{\bibfnamefont{M.}~\bibnamefont{{Viel}}},
  \bibinfo{author}{\bibfnamefont{M.~G.} \bibnamefont{{Haehnelt}}},
  \bibnamefont{and}
  \bibinfo{author}{\bibfnamefont{V.}~\bibnamefont{{Springel}}},
  \bibinfo{journal}{\mnras} \textbf{\bibinfo{volume}{354}},
  \bibinfo{pages}{684} (\bibinfo{year}{2004}), \eprint{arXiv:astro-ph/0404600}.

\bibitem[{\citenamefont{{Kim} et~al.}(2004)\citenamefont{{Kim}, {Viel},
  {Haehnelt}, {Carswell}, and {Cristiani}}}]{Kim2004}
\bibinfo{author}{\bibfnamefont{T.-S.} \bibnamefont{{Kim}}},
  \bibinfo{author}{\bibfnamefont{M.}~\bibnamefont{{Viel}}},
  \bibinfo{author}{\bibfnamefont{M.~G.} \bibnamefont{{Haehnelt}}},
  \bibinfo{author}{\bibfnamefont{R.~F.} \bibnamefont{{Carswell}}},
  \bibnamefont{and}
  \bibinfo{author}{\bibfnamefont{S.}~\bibnamefont{{Cristiani}}},
  \bibinfo{journal}{\mnras} \textbf{\bibinfo{volume}{347}},
  \bibinfo{pages}{355} (\bibinfo{year}{2004}), \eprint{arXiv:astro-ph/0308103}.

\bibitem[{\citenamefont{{Croft} et~al.}(2002)\citenamefont{{Croft}, {Weinberg},
  {Bolte}, {Burles}, {Hernquist}, {Katz}, {Kirkman}, and {Tytler}}}]{Croft2002}
\bibinfo{author}{\bibfnamefont{R.~A.~C.} \bibnamefont{{Croft}}},
  \bibinfo{author}{\bibfnamefont{D.~H.} \bibnamefont{{Weinberg}}},
  \bibinfo{author}{\bibfnamefont{M.}~\bibnamefont{{Bolte}}},
  \bibinfo{author}{\bibfnamefont{S.}~\bibnamefont{{Burles}}},
  \bibinfo{author}{\bibfnamefont{L.}~\bibnamefont{{Hernquist}}},
  \bibinfo{author}{\bibfnamefont{N.}~\bibnamefont{{Katz}}},
  \bibinfo{author}{\bibfnamefont{D.}~\bibnamefont{{Kirkman}}},
  \bibnamefont{and} \bibinfo{author}{\bibfnamefont{D.}~\bibnamefont{{Tytler}}},
  \bibinfo{journal}{\apj} \textbf{\bibinfo{volume}{581}}, \bibinfo{pages}{20}
  (\bibinfo{year}{2002}), \eprint{arXiv:astro-ph/0012324}.

\bibitem[{\citenamefont{{McDonald} et~al.}(2005)\citenamefont{{McDonald},
  {Seljak}, {Cen}, {Shih}, {Weinberg}, {Burles}, {Schneider}, {Schlegel},
  {Bahcall}, {Briggs} et~al.}}]{McDonald2005}
\bibinfo{author}{\bibfnamefont{P.}~\bibnamefont{{McDonald}}},
  \bibinfo{author}{\bibfnamefont{U.}~\bibnamefont{{Seljak}}},
  \bibinfo{author}{\bibfnamefont{R.}~\bibnamefont{{Cen}}},
  \bibinfo{author}{\bibfnamefont{D.}~\bibnamefont{{Shih}}},
  \bibinfo{author}{\bibfnamefont{D.~H.} \bibnamefont{{Weinberg}}},
  \bibinfo{author}{\bibfnamefont{S.}~\bibnamefont{{Burles}}},
  \bibinfo{author}{\bibfnamefont{D.~P.} \bibnamefont{{Schneider}}},
  \bibinfo{author}{\bibfnamefont{D.~J.} \bibnamefont{{Schlegel}}},
  \bibinfo{author}{\bibfnamefont{N.~A.} \bibnamefont{{Bahcall}}},
  \bibinfo{author}{\bibfnamefont{J.~W.} \bibnamefont{{Briggs}}},
  \bibnamefont{et~al.}, \bibinfo{journal}{\apj} \textbf{\bibinfo{volume}{635}},
  \bibinfo{pages}{761} (\bibinfo{year}{2005}), \eprint{arXiv:astro-ph/0407377}.

\bibitem[{\citenamefont{{McDonald} et~al.}(2006)\citenamefont{{McDonald},
  {Seljak}, {Burles}, {Schlegel}, {Weinberg}, {Cen}, {Shih}, {Schaye},
  {Schneider}, {Bahcall} et~al.}}]{McDonald2006}
\bibinfo{author}{\bibfnamefont{P.}~\bibnamefont{{McDonald}}},
  \bibinfo{author}{\bibfnamefont{U.}~\bibnamefont{{Seljak}}},
  \bibinfo{author}{\bibfnamefont{S.}~\bibnamefont{{Burles}}},
  \bibinfo{author}{\bibfnamefont{D.~J.} \bibnamefont{{Schlegel}}},
  \bibinfo{author}{\bibfnamefont{D.~H.} \bibnamefont{{Weinberg}}},
  \bibinfo{author}{\bibfnamefont{R.}~\bibnamefont{{Cen}}},
  \bibinfo{author}{\bibfnamefont{D.}~\bibnamefont{{Shih}}},
  \bibinfo{author}{\bibfnamefont{J.}~\bibnamefont{{Schaye}}},
  \bibinfo{author}{\bibfnamefont{D.~P.} \bibnamefont{{Schneider}}},
  \bibinfo{author}{\bibfnamefont{N.~A.} \bibnamefont{{Bahcall}}},
  \bibnamefont{et~al.}, \bibinfo{journal}{\apjs}
  \textbf{\bibinfo{volume}{163}}, \bibinfo{pages}{80} (\bibinfo{year}{2006}),
  \eprint{arXiv:astro-ph/0405013}.

\bibitem[{\citenamefont{Dimopoulos et~al.}(1996)\citenamefont{Dimopoulos,
  Giudice, and Pomarol}}]{Dimopoulos:1996gy}
\bibinfo{author}{\bibfnamefont{S.}~\bibnamefont{Dimopoulos}},
  \bibinfo{author}{\bibfnamefont{G.~F.} \bibnamefont{Giudice}},
  \bibnamefont{and} \bibinfo{author}{\bibfnamefont{A.}~\bibnamefont{Pomarol}},
  \bibinfo{journal}{Phys. Lett.} \textbf{\bibinfo{volume}{B389}},
  \bibinfo{pages}{37} (\bibinfo{year}{1996}), \eprint{hep-ph/9607225}.

\bibitem[{\citenamefont{Chamseddine et~al.}(1982)\citenamefont{Chamseddine,
  Arnowitt, and Nath}}]{Chamseddine:1982jx}
\bibinfo{author}{\bibfnamefont{A.~H.} \bibnamefont{Chamseddine}},
  \bibinfo{author}{\bibfnamefont{R.~L.} \bibnamefont{Arnowitt}},
  \bibnamefont{and} \bibinfo{author}{\bibfnamefont{P.}~\bibnamefont{Nath}},
  \bibinfo{journal}{Phys. Rev. Lett.} \textbf{\bibinfo{volume}{49}},
  \bibinfo{pages}{970} (\bibinfo{year}{1982}).

\bibitem[{\citenamefont{Barbieri et~al.}(1982)\citenamefont{Barbieri, Ferrara,
  and Savoy}}]{Barbieri:1982eh}
\bibinfo{author}{\bibfnamefont{R.}~\bibnamefont{Barbieri}},
  \bibinfo{author}{\bibfnamefont{S.}~\bibnamefont{Ferrara}}, \bibnamefont{and}
  \bibinfo{author}{\bibfnamefont{C.~A.} \bibnamefont{Savoy}},
  \bibinfo{journal}{Phys. Lett.} \textbf{\bibinfo{volume}{B119}},
  \bibinfo{pages}{343} (\bibinfo{year}{1982}).

\bibitem[{\citenamefont{Hall et~al.}(1983)\citenamefont{Hall, Lykken, and
  Weinberg}}]{Hall:1983iz}
\bibinfo{author}{\bibfnamefont{L.~J.} \bibnamefont{Hall}},
  \bibinfo{author}{\bibfnamefont{J.~D.} \bibnamefont{Lykken}},
  \bibnamefont{and} \bibinfo{author}{\bibfnamefont{S.}~\bibnamefont{Weinberg}},
  \bibinfo{journal}{Phys. Rev.} \textbf{\bibinfo{volume}{D27}},
  \bibinfo{pages}{2359} (\bibinfo{year}{1983}).

\bibitem[{\citenamefont{Cremmer et~al.}(1983)\citenamefont{Cremmer, Fayet, and
  Girardello}}]{Cremmer:1982vy}
\bibinfo{author}{\bibfnamefont{E.}~\bibnamefont{Cremmer}},
  \bibinfo{author}{\bibfnamefont{P.}~\bibnamefont{Fayet}}, \bibnamefont{and}
  \bibinfo{author}{\bibfnamefont{L.}~\bibnamefont{Girardello}},
  \bibinfo{journal}{Phys. Lett.} \textbf{\bibinfo{volume}{B122}},
  \bibinfo{pages}{41} (\bibinfo{year}{1983}).

\bibitem[{\citenamefont{Ohta}(1983)}]{Ohta:1982wn}
\bibinfo{author}{\bibfnamefont{N.}~\bibnamefont{Ohta}}, \bibinfo{journal}{Prog.
  Theor. Phys.} \textbf{\bibinfo{volume}{70}}, \bibinfo{pages}{542}
  (\bibinfo{year}{1983}).

\bibitem[{\citenamefont{Goh et~al.}(2005)\citenamefont{Goh, Luty, and
  Ng}}]{Goh:2003yr}
\bibinfo{author}{\bibfnamefont{H.-S.} \bibnamefont{Goh}},
  \bibinfo{author}{\bibfnamefont{M.~A.} \bibnamefont{Luty}}, \bibnamefont{and}
  \bibinfo{author}{\bibfnamefont{S.-P.} \bibnamefont{Ng}},
  \bibinfo{journal}{JHEP} \textbf{\bibinfo{volume}{01}}, \bibinfo{pages}{040}
  (\bibinfo{year}{2005}), \eprint{hep-th/0309103}.

\bibitem[{\citenamefont{Goh et~al.}(2006)\citenamefont{Goh, Ng, and
  Okada}}]{Goh:2005be}
\bibinfo{author}{\bibfnamefont{H.-S.} \bibnamefont{Goh}},
  \bibinfo{author}{\bibfnamefont{S.-P.} \bibnamefont{Ng}}, \bibnamefont{and}
  \bibinfo{author}{\bibfnamefont{N.}~\bibnamefont{Okada}},
  \bibinfo{journal}{JHEP} \textbf{\bibinfo{volume}{01}}, \bibinfo{pages}{147}
  (\bibinfo{year}{2006}), \eprint{hep-ph/0511301}.

\bibitem[{\citenamefont{Ng and Okada}(2007)}]{Ng:2007sm}
\bibinfo{author}{\bibfnamefont{S.-P.} \bibnamefont{Ng}} \bibnamefont{and}
  \bibinfo{author}{\bibfnamefont{N.}~\bibnamefont{Okada}},
  \bibinfo{journal}{JHEP} \textbf{\bibinfo{volume}{09}}, \bibinfo{pages}{040}
  (\bibinfo{year}{2007}), \eprint{0705.2258}.

\bibitem[{\citenamefont{Dawson}(1985)}]{Dawson:1985vr}
\bibinfo{author}{\bibfnamefont{S.}~\bibnamefont{Dawson}},
  \bibinfo{journal}{Nucl. Phys.} \textbf{\bibinfo{volume}{B261}},
  \bibinfo{pages}{297} (\bibinfo{year}{1985}).

\bibitem[{\citenamefont{Dreiner and Morawitz}(1994)}]{Dreiner:1994tj}
\bibinfo{author}{\bibfnamefont{H.~K.} \bibnamefont{Dreiner}} \bibnamefont{and}
  \bibinfo{author}{\bibfnamefont{P.}~\bibnamefont{Morawitz}},
  \bibinfo{journal}{Nucl. Phys.} \textbf{\bibinfo{volume}{B428}},
  \bibinfo{pages}{31} (\bibinfo{year}{1994}), \eprint{hep-ph/9405253}.

\bibitem[{\citenamefont{Baltz and Gondolo}(1998)}]{Baltz:1997gd}
\bibinfo{author}{\bibfnamefont{E.~A.} \bibnamefont{Baltz}} \bibnamefont{and}
  \bibinfo{author}{\bibfnamefont{P.}~\bibnamefont{Gondolo}},
  \bibinfo{journal}{Phys. Rev.} \textbf{\bibinfo{volume}{D57}},
  \bibinfo{pages}{2969} (\bibinfo{year}{1998}), \eprint{hep-ph/9709445}.

\bibitem[{\citenamefont{Chemtob}(2005)}]{Chemtob:2004xr}
\bibinfo{author}{\bibfnamefont{M.}~\bibnamefont{Chemtob}},
  \bibinfo{journal}{Prog. Part. Nucl. Phys.} \textbf{\bibinfo{volume}{54}},
  \bibinfo{pages}{71} (\bibinfo{year}{2005}), \eprint{hep-ph/0406029}.

\bibitem[{\citenamefont{Landau}(1948)}]{Landau:1948}
\bibinfo{author}{\bibfnamefont{L.~D.} \bibnamefont{Landau}},
  \bibinfo{journal}{Dokl. Akad. Nauk Ser. Fiz.} \textbf{\bibinfo{volume}{60}},
  \bibinfo{pages}{207} (\bibinfo{year}{1948}).

\bibitem[{\citenamefont{Yang}(1950)}]{Yang:1950rg}
\bibinfo{author}{\bibfnamefont{C.-N.} \bibnamefont{Yang}},
  \bibinfo{journal}{Phys. Rev.} \textbf{\bibinfo{volume}{77}},
  \bibinfo{pages}{242} (\bibinfo{year}{1950}).

\bibitem[{\citenamefont{Holdom}(1986)}]{Holdom:1985ag}
\bibinfo{author}{\bibfnamefont{B.}~\bibnamefont{Holdom}},
  \bibinfo{journal}{Phys. Lett.} \textbf{\bibinfo{volume}{B166}},
  \bibinfo{pages}{196} (\bibinfo{year}{1986}).

\bibitem[{\citenamefont{Babu et~al.}(1996)\citenamefont{Babu, Kolda, and
  March-Russell}}]{Babu:1996vt}
\bibinfo{author}{\bibfnamefont{K.~S.} \bibnamefont{Babu}},
  \bibinfo{author}{\bibfnamefont{C.~F.} \bibnamefont{Kolda}}, \bibnamefont{and}
  \bibinfo{author}{\bibfnamefont{J.}~\bibnamefont{March-Russell}},
  \bibinfo{journal}{Phys. Rev.} \textbf{\bibinfo{volume}{D54}},
  \bibinfo{pages}{4635} (\bibinfo{year}{1996}), \eprint{hep-ph/9603212}.

\bibitem[{\citenamefont{Babu et~al.}(1998)\citenamefont{Babu, Kolda, and
  March-Russell}}]{Babu:1997st}
\bibinfo{author}{\bibfnamefont{K.~S.} \bibnamefont{Babu}},
  \bibinfo{author}{\bibfnamefont{C.~F.} \bibnamefont{Kolda}}, \bibnamefont{and}
  \bibinfo{author}{\bibfnamefont{J.}~\bibnamefont{March-Russell}},
  \bibinfo{journal}{Phys. Rev.} \textbf{\bibinfo{volume}{D57}},
  \bibinfo{pages}{6788} (\bibinfo{year}{1998}), \eprint{hep-ph/9710441}.

\bibitem[{\citenamefont{Cheng and Low}(2004)}]{Cheng:2004yc}
\bibinfo{author}{\bibfnamefont{H.-C.} \bibnamefont{Cheng}} \bibnamefont{and}
  \bibinfo{author}{\bibfnamefont{I.}~\bibnamefont{Low}},
  \bibinfo{journal}{JHEP} \textbf{\bibinfo{volume}{08}}, \bibinfo{pages}{061}
  (\bibinfo{year}{2004}), \eprint{hep-ph/0405243}.

\bibitem[{\citenamefont{Csaki et~al.}(2008)\citenamefont{Csaki, Heinonen,
  Perelstein, and Spethmann}}]{Csaki:2008se}
\bibinfo{author}{\bibfnamefont{C.}~\bibnamefont{Csaki}},
  \bibinfo{author}{\bibfnamefont{J.}~\bibnamefont{Heinonen}},
  \bibinfo{author}{\bibfnamefont{M.}~\bibnamefont{Perelstein}},
  \bibnamefont{and} \bibinfo{author}{\bibfnamefont{C.}~\bibnamefont{Spethmann}}
  (\bibinfo{year}{2008}), \eprint{0804.0622}.

\bibitem[{\citenamefont{Krohn and Yavin}(2008)}]{Krohn:2008ye}
\bibinfo{author}{\bibfnamefont{D.}~\bibnamefont{Krohn}} \bibnamefont{and}
  \bibinfo{author}{\bibfnamefont{I.}~\bibnamefont{Yavin}},
  \bibinfo{journal}{JHEP} \textbf{\bibinfo{volume}{06}}, \bibinfo{pages}{092}
  (\bibinfo{year}{2008}), \eprint{0803.4202}.

\bibitem[{\citenamefont{Barger et~al.}(2007)\citenamefont{Barger, Keung, and
  Gao}}]{Barger:2007df}
\bibinfo{author}{\bibfnamefont{V.}~\bibnamefont{Barger}},
  \bibinfo{author}{\bibfnamefont{W.-Y.} \bibnamefont{Keung}}, \bibnamefont{and}
  \bibinfo{author}{\bibfnamefont{Y.}~\bibnamefont{Gao}},
  \bibinfo{journal}{Phys. Lett.} \textbf{\bibinfo{volume}{B655}},
  \bibinfo{pages}{228} (\bibinfo{year}{2007}), \eprint{0707.3648}.

\bibitem[{\citenamefont{Freitas et~al.}(2008)\citenamefont{Freitas, Schwaller,
  and Wyler}}]{Freitas:2008mq}
\bibinfo{author}{\bibfnamefont{A.}~\bibnamefont{Freitas}},
  \bibinfo{author}{\bibfnamefont{P.}~\bibnamefont{Schwaller}},
  \bibnamefont{and} \bibinfo{author}{\bibfnamefont{D.}~\bibnamefont{Wyler}}
  (\bibinfo{year}{2008}), \eprint{0806.3674}.

\bibitem[{\citenamefont{Strong et~al.}(2005)}]{Strong:2005zx}
\bibinfo{author}{\bibfnamefont{A.~W.} \bibnamefont{Strong}}
  \bibnamefont{et~al.}, \bibinfo{journal}{Astron. Astrophys.}
  \textbf{\bibinfo{volume}{444}}, \bibinfo{pages}{495} (\bibinfo{year}{2005}),
  \eprint{astro-ph/0509290}.

\bibitem[{\citenamefont{Aguilar et~al.}(2007)}]{Aguilar:2007yf}
\bibinfo{author}{\bibfnamefont{M.}~\bibnamefont{Aguilar}} \bibnamefont{et~al.}
  (\bibinfo{collaboration}{AMS-01}), \bibinfo{journal}{Phys. Lett.}
  \textbf{\bibinfo{volume}{B646}}, \bibinfo{pages}{145} (\bibinfo{year}{2007}),
  \eprint{astro-ph/0703154}.

\bibitem[{\citenamefont{Bernabei et~al.}(2008)}]{Bernabei:2008yi}
\bibinfo{author}{\bibfnamefont{R.}~\bibnamefont{Bernabei}} \bibnamefont{et~al.}
  (\bibinfo{collaboration}{DAMA}), \bibinfo{journal}{Eur. Phys. J.}
  \textbf{\bibinfo{volume}{C56}}, \bibinfo{pages}{333} (\bibinfo{year}{2008}),
  \eprint{0804.2741}.

\bibitem[{\citenamefont{Adriani et~al.}(2008)}]{Adriani:2008zr}
\bibinfo{author}{\bibfnamefont{O.}~\bibnamefont{Adriani}} \bibnamefont{et~al.}
  (\bibinfo{year}{2008}), \eprint{0810.4995}.

\end{thebibliography}

\end{document}